\documentclass[pdftex,twocolumn,epjc3_preprint,runningheads]{svjour3}

\usepackage[T1]{fontenc}
\usepackage{lmodern}
\usepackage{calc}
\usepackage{graphicx}
\usepackage{booktabs}
\usepackage{textcomp}
\usepackage{xspace}
\usepackage{relsize}
\usepackage{amssymb}
\usepackage{amsmath}
\usepackage{listings}
\usepackage{microtype}
\usepackage{multirow}
\usepackage{tabularx}
\usepackage{array}
\usepackage{placeins}
\usepackage{cuted}
\usepackage{soul} 
\usepackage{fixltx2e}
\usepackage[numbers,sort&compress]{natbib}
\usepackage[labelfont=bf,font=small]{caption}
\usepackage[skip=-2pt]{subcaption}
\usepackage[colorlinks,citecolor=blue,urlcolor=blue,linkcolor=blue,breaklinks=breakall]{hyperref}
\usepackage{breakurl}
\usepackage[dvipsnames]{xcolor}
\usepackage[clockwise,figuresright]{rotating}
\usepackage{siunitx}
\usepackage{tikz}

\usepackage{etoolbox}
\AfterEndEnvironment{strip}{\leavevmode}

\allowdisplaybreaks

\newcolumntype{L}{>{\raggedright\let\newline\\\arraybackslash\hspace{0pt}}X}
\newcolumntype{R}{>{\raggedleft\let\newline\\\arraybackslash\hspace{0pt}}X}
\newcolumntype{C}{>{\centering\let\newline\\\arraybackslash\hspace{0pt}}X}

\newcommand{\imperial}{Department of Physics, Imperial College London, Blackett Laboratory, Prince Consort Road, London SW7 2AZ, UK}
\newcommand{\nordita}{NORDITA, Roslagstullsbacken 23, SE-10691 Stockholm, Sweden}
\newcommand{\oslo}{Department of Physics, University of Oslo, N-0316 Oslo, Norway}
\newcommand{\adelaide}{Department of Physics, University of Adelaide, Adelaide, SA 5005, Australia}
\newcommand{\glasgow}{SUPA, School of Physics and Astronomy, University of Glasgow, Glasgow, G12 8QQ, UK}
\newcommand{\monash}{School of Physics and Astronomy, Monash University, Melbourne, VIC 3800, Australia}
\newcommand{\coepp}{Australian Research Council Centre of Excellence for Particle Physics at the Tera-scale}
\newcommand{\okc}{Oskar Klein Centre for Cosmoparticle Physics, AlbaNova University Centre, SE-10691 Stockholm, Sweden}
\newcommand{\su}{Department of Physics, Stockholm University, SE-10691 Stockholm, Sweden}
\newcommand{\mcgill}{Department of Physics, McGill University, 3600 rue University, Montr\'eal, Qu\'ebec H3A 2T8, Canada}
\newcommand{\ucla}{Physics and Astronomy Department, University of California, Los Angeles, CA 90095, USA}
\newcommand{\annecy}{LAPTh, Universit\'e de Savoie, CNRS, 9 chemin de Bellevue B.P.110, F-74941 Annecy-le-Vieux, France}
\newcommand{\harvard}{Department of Physics, Harvard University, Cambridge, MA 02138, USA}
\newcommand{\grappa}{GRAPPA, Institute of Physics, University of Amsterdam, Science Park 904, 1098 XH Amsterdam, Netherlands}
\newcommand{\sydney}{Centre for Translational Data Science, Faculty of Engineering and Information Technologies, School of Physics, The University of Sydney, NSW 2006, Australia}

\newcommand{\desy}{DESY, Notkestra\ss e 85, D-22607 Hamburg, Germany}

\newcommand{\cernth}{Theoretical Physics Department, CERN, CH-1211 Geneva 23, Switzerland}
\newcommand{\lyon}{Univ Lyon, Univ Lyon 1, ENS de Lyon, CNRS, Centre de Recherche Astrophysique de Lyon UMR5574, F-69230 Saint-Genis-Laval, France}
\newcommand{\iuf}{Institut Universitaire de France, 103 boulevard Saint-Michel, 75005 Paris, France}
\newcommand{\zurich}{Physik-Institut, Universit\"at Z\"urich, Winterthurerstrasse 190, 8057 Z\"urich, Switzerland}
\newcommand{\krakow}{H.~Niewodnicza\'nski Institute of Nuclear Physics, Polish Academy of Sciences, 31-342  Krak\'ow, Poland}

\newcommand{\gambitacknos    }{We warmly thank the Casa Matem\'aticas Oaxaca, affiliated with the Banff International Research Station, for hospitality whilst part of this work was completed, and the staff at Cyfronet, for their always helpful supercomputing support.  \GB has been supported by STFC (UK; ST/K00414X/1, ST/P000762/1), the Royal Society (UK; UF110191), Glasgow University (UK; Leadership Fellowship), the Research Council of Norway (FRIPRO 230546/F20), NOTUR (Norway; NN9284K), the Knut and Alice Wallenberg Foundation (Sweden; Wallenberg Academy Fellowship), the Swedish Research Council (621-2014-5772), the Australian Research Council (CE110001004, FT130100018, FT140100244, FT160100274), The University of Sydney (Australia; IRCA-G162448), PLGrid Infrastructure (Poland), Polish National Science Center (Sonata UMO-2015/17/D/ST2/03532), the Swiss National Science Foundation (PP00P2-144674), the European Commission Horizon 2020 Marie Sk\l{}odowska-Curie actions (H2020-MSCA-RISE-2015-691164), the ERA-CAN+ Twinning Program (EU \& Canada), the Netherlands Organisation for Scientific Research (NWO-Vidi 680-47-532), the National Science Foundation (USA; DGE-1339067), the FRQNT (Qu\'ebec) and NSERC/The Canadian Tri-Agencies Research Councils (BPDF-424460-2012).}

\makeatletter

\newcommand{\preprintnumber}[1]{\gdef\@preprintnumber{\begin{flushright}{#1}\end{flushright}}}

\g@addto@macro\bfseries{\boldmath}
\makeatother

\newcommand{\subparagraph}{} 
\usepackage{titlesec}
\titleformat*{\paragraph}{\bfseries}

\journalname{Eur. Phys. J. C}
\smartqed
\let\underscore\_
\renewcommand{\_}{\discretionary{\underscore}{}{\underscore}}

\makeatletter
\let\orgdescriptionlabel\descriptionlabel
\renewcommand*{\descriptionlabel}[1]{%
  \let\orglabel\label
  \let\label\@gobble
  \phantomsection
  \protected@edef\@currentlabel{#1}%
  \let\label\orglabel
  \orgdescriptionlabel{#1}%
}
\makeatother

\newcommand\postnewlinemarker{\hbox{\ensuremath{\hookrightarrow}}}
\newcommand\cpp[1]{{\lstinline!#1!}}  

\newcommand\yaml[1]{{\lstset{style=yaml}\lstinline!#1!\lstset{style=cpp}}}

\newcommand\term[1]{{\lstset{style=terminal}\lstinline!#1!\lstset{style=cpp}}}
\newcommand\fortran[1]{{\lstset{style=fortran}\lstinline!#1!\lstset{style=cpp}}}
\newcommand\py[1]{{\lstset{style=python}\lstinline!#1!\lstset{style=cpp}}}
\newcommand\customtilde{{\raisebox{0.2ex}{\scalebox{0.6}{\boldmath$\sim$}}}}
\newcommand\mathematica[1]{{\lstset{style=Mathematica}\lstinline!#1!\lstset{style=cpp}}}

\lstnewenvironment{lstlistingyaml}{\lstset{style=yaml}}{\lstset{style=cpp}}
\lstnewenvironment{lstlistingterm}{\lstset{style=terminal}}{\lstset{style=cpp}}
\lstnewenvironment{lstlistingfortran}{\lstset{style=fortran}}{\lstset{style=cpp}}
\lstnewenvironment{lstcpp}{\lstset{style=cpp}}{\lstset{style=cpp}}
\lstnewenvironment{lstcppalt}{\lstset{style=cppalt}}{\lstset{style=cpp}}
\lstnewenvironment{lstcppnum}{\lstset{style=cppnum}}{\lstset{style=cpp}}
\lstnewenvironment{lstyaml}{\lstset{style=yaml}}{\lstset{style=cpp}}
\lstnewenvironment{lstterm}{\lstset{style=terminal}}{\lstset{style=cpp}}
\lstnewenvironment{lsttermalt}{\lstset{style=terminalalt}}{\lstset{style=cpp}}
\lstnewenvironment{lsttext}{\lstset{style=text}}{\lstset{style=cpp}}
\lstnewenvironment{lstfortran}{\lstset{style=fortran}}{\lstset{style=cpp}}
\lstnewenvironment{lstpy}{\lstset{style=python}}{\lstset{style=cpp}}

\newcommand{\tmpname}{}
\newcommand{\tmplistingname}{}
\makeatletter
\newif\ifATOlabelname
\lst@Key{labelname}{Listing}{\def\ATOlabelname{#1}\global\ATOlabelnametrue}
\makeatother
\lstnewenvironment{lstcpplabel}[1][]{
  \lstset{style=cpp,#1} 
  \ifATOlabelname
    \renewcommand{\tmpname}{\lstlistingname}
    \renewcommand{\tmplistingname}{\lstlistlistingname}
    \renewcommand{\lstlistingname}{\ATOlabelname}
    \renewcommand{\lstlistlistingname}{List of \lstlistingname s}
  \fi
}{
  \renewcommand{\lstlistingname}{\tmpname}
  \renewcommand{\lstlistlistingname}{\tmplistingname}
  \lstset{style=cpp}
}
\definecolor{solarized@base03}{HTML}{002B36}
\definecolor{solarized@base02}{HTML}{073642}
\definecolor{solarized@base01}{HTML}{586e75}
\definecolor{solarized@base00}{HTML}{657b83}
\definecolor{solarized@base0}{HTML}{839496}
\definecolor{solarized@base1}{HTML}{93a1a1}
\definecolor{solarized@base2}{HTML}{EEE8D5}
\definecolor{solarized@base3}{HTML}{FDF6E3}
\definecolor{solarized@yellow}{HTML}{B58900}
\definecolor{solarized@orange}{HTML}{CB4B16}
\definecolor{solarized@red}{HTML}{DC322F}
\definecolor{solarized@magenta}{HTML}{D33682}
\definecolor{solarized@violet}{HTML}{6C71C4}
\definecolor{solarized@blue}{HTML}{268BD2}
\definecolor{solarized@cyan}{HTML}{2AA198}
\definecolor{solarized@green}{HTML}{859900}
\definecolor{darkred}{HTML}{550003}
\definecolor{darkgreen}{HTML}{00AA00}

\newcommand\YAMLstringstyle{\footnotesize\color{solarized@green}\mdseries}
\newcommand\YAMLkeystyle{\footnotesize\color{solarized@blue}\ttfamily}
\newcommand\YAMLvaluestyle{\footnotesize\color{blue}\mdseries}
\newcommand\ProcessThreeDashes{\llap{\color{cyan}\mdseries-{-}-}}

\newcommand\CPPcommentstyle{\color{solarized@violet}\footnotesize\ttfamily}
\newcommand\CPPdirectivestyle{\color{solarized@magenta}\footnotesize\ttfamily}
\newcommand\termplainstyle{\footnotesize\ttfamily}

\newcommand\processLongMacroDelimiter
{%
\CPPdirectivestyle%
\#define%
}

\lstdefinestyle{cpp}
{
  language=C++,
  basicstyle=\footnotesize\ttfamily,
  basewidth={0.53em,0.44em}, 
  numbers=none,
  tabsize=2,
  breaklines=true,
  escapeinside={@}{@},
  showstringspaces=false,
  numberstyle=\tiny\color{solarized@base01},
  keywordstyle=\color{solarized@orange},
  stringstyle=\color{solarized@red}\ttfamily,
  identifierstyle=\color{solarized@blue},
  commentstyle=\CPPcommentstyle,
  directivestyle=\CPPdirectivestyle,
  emphstyle=\color{solarized@green},
  frame=single,
  rulecolor=\color{solarized@base2},
  rulesepcolor=\color{solarized@base2},
  literate={~} {\customtilde}1,
  moredelim=*[directive]\ \ \#,
  moredelim=*[directive]\ \ \ \ \#
}

\lstdefinestyle{cppalt}
{
  language=C++,
  basicstyle=\footnotesize\ttfamily,
  basewidth={0.53em,0.44em}, 
  numbers=none,
  tabsize=2,
  breaklines=true,
  escapeinside={*@}{@*},
  showstringspaces=false,
  numberstyle=\tiny\color{solarized@base01},
  keywordstyle=\color{solarized@orange},
  stringstyle=\color{solarized@red}\ttfamily,
  identifierstyle=\color{solarized@blue},
  commentstyle=\CPPcommentstyle,
  directivestyle=\CPPdirectivestyle,
  emphstyle=\color{solarized@green},
  frame=single,
  rulecolor=\color{solarized@base2},
  rulesepcolor=\color{solarized@base2},
  literate={~}{\customtilde}1,
  moredelim=**[is][\processLongMacroDelimiter]{BeginLongMacro}{EndLongMacro} 
}

\lstdefinestyle{cppnum}
{
  language=C++,
  basicstyle=\footnotesize\ttfamily,
  basewidth={0.53em,0.44em}, 
  numbers=none,
  tabsize=2,
  breaklines=true,
  escapeinside={@}{@},
  numberstyle=\tiny\color{solarized@base01},
  showstringspaces=false,
  numberstyle=\tiny\color{solarized@base01},
  keywordstyle=\color{solarized@orange},
  stringstyle=\color{solarized@red}\ttfamily,
  identifierstyle=\color{solarized@blue},
  commentstyle=\CPPcommentstyle,
  directivestyle=\CPPdirectivestyle,
  emphstyle=\color{solarized@green},
  frame=single,
  rulecolor=\color{solarized@base2},
  rulesepcolor=\color{solarized@base2},
  literate={~} {\customtilde}1,
  moredelim=*[directive]\ \ \#,
  moredelim=*[directive]\ \ \ \ \#
}

\lstdefinestyle{python}
{
  language=Python,
  basicstyle=\footnotesize\ttfamily,
  basewidth={0.53em,0.44em},
  numbers=none,
  tabsize=2,
  breaklines=true,
  escapeinside={@}{@},
  showstringspaces=false,
  numberstyle=\tiny\color{solarized@base01},
  keywordstyle=\color{blue},
  stringstyle=\color{orange}\ttfamily,
  identifierstyle=\color{darkred},
  commentstyle=\color{purple},
  emphstyle=\color{green},
  frame=single,
  rulecolor=\color{solarized@base2},
  rulesepcolor=\color{solarized@base2},
  literate = {~}{\customtilde}1
             {\ as\ }{{\color{blue}\ as\ \color{black}}}3
}

\lstdefinestyle{fortran}
{
  language=Fortran,
  basicstyle=\footnotesize\ttfamily,
  basewidth={0.53em,0.44em},
  numbers=none,
  tabsize=2,
  breaklines=true,
  escapeinside={@}{@},
  showstringspaces=false,
  numberstyle=\tiny\color{solarized@base01},
  keywordstyle=\color{blue},
  stringstyle=\color{orange}\ttfamily,
  identifierstyle=\color{Periwinkle},
  commentstyle=\color{purple},
  emphstyle=\color{green},
  morekeywords={and, or, true, false},
  frame=single,
  rulecolor=\color{solarized@base2},
  rulesepcolor=\color{solarized@base2},
  literate={~}{\customtilde}1
}

\lstdefinestyle{terminal}
{
  language=bash,
  basicstyle=\termplainstyle,
  numbers=none,
  tabsize=2,
  breaklines=true,
  escapeinside={@}{@},
  frame=single,
  showstringspaces=false,
  numberstyle=\tiny\color{solarized@base01},
  keywordstyle=\color{solarized@orange},
  stringstyle=\color{solarized@red}\ttfamily,
  identifierstyle=\color{black},
  commentstyle=\color{solarized@violet},
  emphstyle=\color{solarized@green},
  frame=single,
  rulecolor=\color{solarized@base2},
  rulesepcolor=\color{solarized@base2},
  morekeywords={gambit, cmake, make, mkdir},
  deletekeywords={test},
  literate = {\ gambit}{{\ }{\color{black}}gambit}7
             {/gambit}{{/}{\color{black}}gambit}6
             {gambit/}{{\color{black}}gambit{/}}6
             {/include}{{/}{\color{black}}include}8
             {cmake/}{{\color{black}}cmake/}6
             {.cmake}{{.}{\color{black}}cmake}6
             {~}{\customtilde}1
}

\lstdefinestyle{terminalalt}
{
  language=bash,
  basicstyle=\footnotesize\ttfamily,
  numbers=none,
  tabsize=2,
  breaklines=true,
  escapeinside={*@}{@*},
  frame=single,
  showstringspaces=false,
  numberstyle=\tiny\color{solarized@base01},
  keywordstyle=\color{solarized@orange},
  stringstyle=\color{solarized@red}\ttfamily,
  identifierstyle=\color{black},
  commentstyle=\color{solarized@violet},
  emphstyle=\color{solarized@green},
  frame=single,
  rulecolor=\color{solarized@base2},
  rulesepcolor=\color{solarized@base2},
  morekeywords={gambit, cmake, make, mkdir},
  deletekeywords={test},
  literate = {\ gambit}{{\ }{\color{black}}gambit}7
             {/gambit}{{/}{\color{black}}gambit}6
             {gambit/}{{\color{black}}gambit{/}}6
             {/include}{{/}{\color{black}}include}8
             {cmake/}{{\color{black}}cmake/}6
             {.cmake}{{.}{\color{black}}cmake}6
             {~}{\customtilde}1
}

\lstdefinestyle{text}
{
  language={},
  basicstyle=\footnotesize\ttfamily,
  identifierstyle=\color{black},
  numbers=none,
  tabsize=2,
  breaklines=true,
  escapeinside={*@}{@*},
  showstringspaces=false,
  frame=single,
  rulecolor=\color{solarized@base2},
  rulesepcolor=\color{solarized@base2},
  literate={~}{\customtilde}1
}

\lstdefinestyle{yaml}
{
  language=bash,
  escapeinside={@}{@},
  keywords={true,false,null},
  otherkeywords={},
  keywordstyle=\color{solarized@base0}\bfseries,
  basicstyle=\footnotesize\color{black}\ttfamily,
  identifierstyle=\YAMLkeystyle,
  sensitive=false,
  commentstyle=\color{solarized@orange}\ttfamily,
  morecomment=[l]{\#},
  morecomment=[s]{/*}{*/},
  stringstyle=\YAMLstringstyle\ttfamily,
  moredelim=**[s][\YAMLkeystyle]{,}{:},   
  moredelim=**[l][\YAMLvaluestyle]{:},    
  morestring=[b]',
  morestring=[b]",
  literate =    {---}{{\ProcessThreeDashes}}3
                {>}{{\textcolor{solarized@red}\textgreater}}1
                {|}{{\textcolor{solarized@red}\textbar}}1
                {\ -\ }{{\mdseries\color{black}\ -\ \negmedspace}}3
                {\}}{{{\color{black} \}}}}1
                {\{}{{{\color{black} \{}}}1
                {[}{{{\color{black} [}}}1
                {]}{{{\color{black} ]}}}1
                {~}{\customtilde}1,
  breakindent=0pt,
  breakatwhitespace,
  columns=fullflexible
}

\lstdefinestyle{mathematica}
{
  language={Mathematica},
  basicstyle=\footnotesize\ttfamily,
  basewidth={0.53em,0.44em},
  numbers=none,
  tabsize=2,
  breaklines=true,
  escapeinside={@}{@},
  numberstyle=\tiny\color{black},
  showstringspaces=false,
  numberstyle=\tiny\color{solarized@base01},
  keywordstyle=\color{solarized@orange},
  stringstyle=\color{solarized@red}\ttfamily,
  identifierstyle=\color{solarized@orange}\ttfamily,
  commentstyle=\color{solarized@gray}\ttfamily,
  directivestyle=\color{solarized@orange}\ttfamily,
  emphstyle=\color{solarized@green},
  frame=single,
  rulecolor=\color{solarized@base2},
  rulesepcolor=\color{solarized@base2},
  literate={~} {\customtilde}1,
  moredelim=*[directive]\ \ \#,
  moredelim=*[directive]\ \ \ \ \#,
  mathescape=true
}

\newcommand{\doublecross}[2]{\hyperref[#2]{\textbf{#1}}}
\newcommand{\doublecrosssf}[2]{\hyperref[#2]{\textbf{\textsf{#1}}}}

\newcommand{\startglossary}{\section{Glossary}\label{glossary}Here we explain some terms that have specific technical definitions in \GB.\begin{description}}
\newcommand{\finishglossary}{\end{description}}

\newcommand{\bcode}{\begin{lstlisting}}
\newcommand{\ecode}{\end{lstlisting}}

\newcommand{\sss}{\scriptscriptstyle}
\newcommand{\ms}{m_{\sss S}}
\newcommand{\lhs}{\lambda_{h\sss S}}
\newcommand{\ls}{\lambda_{\sss S}}
\newcommand{\mh}{m_h}

\newcommand{\MSbar}{$\MSBar$\xspace}
\newcommand{\MSBar}{\overline{MS}}

\newcommand{\gambit}{\textsf{GAMBIT}\xspace}

\newcommand{\darkbit}{\textsf{DarkBit}\xspace}

\newcommand{\specbit}{\textsf{SpecBit}\xspace}
\newcommand{\decaybit}{\textsf{DecayBit}\xspace}
\newcommand{\precisionbit}{\textsf{PrecisionBit}\xspace}
\newcommand{\scannerbit}{\textsf{ScannerBit}\xspace}

\newcommand{\GB}{\gambit}

\newcommand{\ds}{\textsf{DarkSUSY}\xspace}
\newcommand{\darksusy}{\ds}

\newcommand\FlexibleSUSY{\textsf{FlexibleSUSY}\xspace}
\newcommand\SOFTSUSY{\textsf{SOFTSUSY}\xspace}

\newcommand\SARAH{\textsf{SARAH}\xspace}

\newcommand\nulike{\textsf{nulike}\xspace}
\newcommand\gamLike{\textsf{gamLike}\xspace}
\newcommand\gamlike{\gamLike}

\newcommand\pippi{\textsf{pippi}\xspace}
\newcommand\MultiNest{\textsf{MultiNest}\xspace}
\newcommand\multinest{\MultiNest}
\newcommand\great{\textsf{GreAT}\xspace}
\newcommand\twalk{\textsf{T-Walk}\xspace}
\newcommand\diver{\textsf{Diver}\xspace}
\newcommand\ddcalc{\textsf{DDCalc}\xspace}

\newcommand\beq{\begin{equation}}
\newcommand\eeq{\end{equation}}

\renewcommand{\url}[1]{\href{#1}{#1}}

\begin{document}

\preprintnumber{CERN-TH-2017-170, CoEPP-MN-17-10, NORDITA 2017-079}

\title{Status of the scalar singlet dark matter model}

\author
{
The GAMBIT Collaboration:
Peter Athron\thanksref{inst:a,inst:b} \and
Csaba Bal\'azs\thanksref{inst:a,inst:b} \and
Torsten Bringmann\thanksref{inst:c} \and
Andy Buckley\thanksref{inst:d} \and
Marcin Chrz\k{a}szcz\thanksref{inst:e,inst:f} \and
Jan Conrad\thanksref{inst:g,inst:h} \and
Jonathan M.~Cornell\thanksref{inst:i,e1} \and
Lars A.~Dal\thanksref{inst:c} \and
Joakim Edsj\"o\thanksref{inst:g,inst:h} \and
Ben Farmer\thanksref{inst:g,inst:h} \and
Paul Jackson\thanksref{inst:k,inst:b} \and
Felix Kahlhoefer\thanksref{inst:v} \and
Abram Krislock\thanksref{inst:c} \and
Anders Kvellestad\thanksref{inst:m} \and
James McKay\thanksref{inst:q,e2} \and
Farvah Mahmoudi\thanksref{inst:n,inst:o,e5} \and
Gregory D.\ Martinez\thanksref{inst:p} \and
Antje Putze\thanksref{inst:r} \and
Are Raklev\thanksref{inst:c} \and
Christopher Rogan\thanksref{inst:s} \and
Aldo Saavedra\thanksref{inst:t,inst:b} \and
Christopher Savage\thanksref{inst:m} \and
Pat Scott\thanksref{inst:q,e3} \and
Nicola Serra\thanksref{inst:e} \and
Christoph Weniger\thanksref{inst:u,e4} \and
Martin White\thanksref{inst:k,inst:b}
}

\institute{%
  \monash\label{inst:a} \and
  \coepp\label{inst:b} \and
  \oslo\label{inst:c} \and
  \glasgow\label{inst:d} \and
  \zurich\label{inst:e} \and
  \krakow\label{inst:f} \and
  \okc\label{inst:g} \and
  \su\label{inst:h} \and
  \mcgill\label{inst:i} \and
  \adelaide\label{inst:k} \and
  \desy\label{inst:v} \and
  \nordita\label{inst:m} \and
  \imperial\label{inst:q} \and
  \lyon\label{inst:n} \and
  \cernth\label{inst:o} \and
  \ucla\label{inst:p} \and
  \annecy\label{inst:r} \and
  \harvard\label{inst:s} \and
  \sydney\label{inst:t} \and
  \grappa\label{inst:u}
}

\thankstext{e1}{cornellj@physics.mcgill.ca}
\thankstext{e2}{j.mckay14@imperial.ac.uk}
\thankstext{e3}{p.scott@imperial.ac.uk}
\thankstext{e4}{c.weniger@uva.nl}
\thankstext[*]{e5}{Also \iuf.}

\titlerunning{Status of the scalar singlet dark matter model}
\authorrunning{The GAMBIT Collaboration}

\date{Received: date / Accepted: date}

\maketitle

\begin{abstract}

One of the simplest viable models for dark matter is an additional neutral scalar, stabilised by a $\mathbb{Z}_2$ symmetry.  Using the \GB package and combining results from four independent samplers, we present Bayesian and frequentist global fits of this model.  We vary the singlet mass and coupling along with 13 nuisance parameters, including nuclear uncertainties relevant for direct detection, the local dark matter density, and selected quark masses and couplings.  We include the dark matter relic density measured by \textit{Planck}, direct searches with LUX, PandaX, SuperCDMS and XENON100, limits on invisible Higgs decays from the Large Hadron Collider, searches for high-energy neutrinos from dark matter annihilation in the Sun with IceCube, and searches for gamma rays from annihilation in dwarf galaxies with the \textit{Fermi}-LAT.  Viable solutions remain at couplings of order unity, for singlet masses between the Higgs mass and about 300\,GeV, and at masses above $\sim$1\,TeV.  Only in the latter case can the scalar singlet constitute all of dark matter.  Frequentist analysis shows that the low-mass resonance region, where the singlet is about half the mass of the Higgs, can also account for all of dark matter, and remains viable.  However, Bayesian considerations show this region to be rather fine-tuned.

\end{abstract}

\tableofcontents

\section{Introduction}
\label{intro}
Dark matter (DM) accounts for the majority of the matter in the Universe, but its nature remains a mystery. It has been known for some time \cite{Bergstrom:2000pn,Bertone:2004pz,Feng:2010gw} that GeV-scale particle DM can accurately reproduce the observed relic abundance of DM, provided that it has an interaction strength with standard model (SM) particles that is comparable to that of the weak force. This is the Weakly Interacting Massive Particle (WIMP) paradigm.

The simplest WIMP model is the ``scalar singlet'' or scalar ``Higgs-portal'' scenario, in which one adds to the SM a massive real scalar field $S$ uncharged under the SM gauge group \cite{SilveiraZee,McDonald94,Burgess01}. $S$ is stabilised by a $\mathbb{Z}_2$ symmetry, and never obtains a vacuum expectation value (VEV). The only renormalisable interactions between the singlet and the SM allowed by the symmetries of the SM arise from a Lagrangian term of the form $S^2H^2$.  This term gives the singlet a so-called ``Higgs-portal'' for interacting with the SM, leading to a range of possible phenomenological consequences.  These include thermal production in the early Universe and present-day annihilation signals \cite{Yaguna09,Profumo2010a,Arina11}, direct detection and $h\to SS$ decays \cite{Mambrini11}.  A number of recent papers have investigated prospects for relaxing these constraints by adding additional scalars \cite{Jana16,Campbell16,Casas17}. The singlet has also been implicated in inflation \cite{Lerner09,Herranen15,Kahlhoefer15} and baryogenesis \cite{Profumo07,Barger09,JimKimmo}.

The simplicity of the scenario and the discovery of the Higgs boson in 2012~\cite{Aad:2012tfa,Chatrchyan2012} have focussed much attention on the singlet model in recent years. XENON100 and WMAP constraints were applied in Ref.\ \cite{Djouadi:2011aa}, and an early global fit of the model using a similar range of data was performed in Ref.~\cite{Cheung:2012xb}.  LHC Run I constraints from a CMS vector boson fusion analysis, and monojet and mono-$Z$ analyses were shown to be very weak \cite{Endo:2014cca}; indeed, monojet constraints on all minimal Higgs-portal models (i.e.\ scalar, fermion or vector DM interacting with the SM only via the Higgs-portal) are weak \cite{Djouadi:2012zc}.  Implications of the Higgs mass measurement and a detailed treatment of direct and indirect detection were explored in Ref.\ \cite{Cline13b}, followed by the application of direct limits from the LUX and PandaX experiments~\cite{Urbano:2014hda,He:2016mls,Escudero:2016gzx}. Anti-proton data can be important in the region of the Higgs resonance \cite{Goudelis09,Urbano:2014hda}, and competitive with the LUX limits at higher DM masses, but are ultimately prone to substantial cosmic ray propagation uncertainties. Discovery prospects at future colliders have been explored for the 14\,TeV LHC and a 100\,TeV hadron collider~\cite{Craig:2014lda,Han:2016gyy}, and the International Linear Collider~\cite{Ko:2016xwd}.

The most comprehensive recent studies were presented in Refs.~\cite{Cline13b,Beniwal} and \cite{Cuoco:2016jqt}. The first pair of papers examined the scalar singlet scenario in light of recent (and projected) LHC Higgs invisible width measurements \cite{Belanger:2013xza,Aad:2015pla,Khachatryan:2016whc}, the \textit{Planck} relic density measurement \cite{Planck15cosmo}, \textit{Planck} and WMAP CMB constraints on DM annihilation at the time of recombination \cite{Cline13,Planck15cosmo,Slatyer15a}, \textit{Fermi}-LAT analysis of gamma rays in the direction of 15 dwarf spheroidal galaxies using 6 years of \texttt{Pass 8} data \cite{LATdwarfP8}, and LUX limits on the spin-independent WIMP-nucleon scattering cross-section \cite{LUX2016}. These studies also investigated the prospects for detection in gamma rays by the Cherenkov Telescope Array (CTA; \cite{Pierre14, Silverwood:2014yza,Carr:2015hta}), and for direct detection by XENON1T \cite{XENON1T}. Ref.~\cite{Cuoco:2016jqt} presented a global fit to determine the regions of the scalar singlet model space that can explain the apparent excess of gamma rays observed by \textit{Fermi} towards the Galactic centre, frequently interpreted as evidence for DM annihilation \cite{Goodenough09, Hooper:2010mq, Hooper:2011ti, Abazajian:2012pn, Gordon:2013vta, Daylan:2014rsa, Calore:2014nla}. This included a treatment of the \textit{Planck} relic density constraint, LHC invisible Higgs width constraints, direct direct search data from LUX (and projections for XENON1T and DARWIN), and constraints from \textit{Fermi}-LAT searches for DM annihilation in dwarf galaxies and $\gamma$-ray lines at the Galactic centre.

Although lines and signals from the Galactic centre in the context of this model have received a reasonable amount of attention \cite{Profumo2010a,Feng15,Duerr15,Duerr16,Cuoco:2016jqt}, in general these signals are relevant only if the singlet is produced non-thermally, as the regions of parameter space where such signals are substantial have quite low thermal relic abundances \cite{Cline13b}.  Fitting the excess at the Galactic centre requires relatively large couplings, which in turn imply too little DM from thermal freeze-out. Some previous studies have solved this issue by assuming an unspecified additional production mechanism.  This reduces the predictability of the theory, as the cosmological abundance of scalar singlets ceases to be a prediction.  We will take a different approach, allowing for the possibility that the scalar singlet constitutes only a sub-component of DM, and permitting a different species (e.g.\ axions) to make up the rest.  Indeed, as we show in this paper, experiments are now so sensitive to DM signals that they can probe singlet models constituting \textit{less than a hundredth of a percent} of the total DM.

The purpose of the present paper is twofold. First and foremost, we provide the most comprehensive study yet of the scalar singlet scenario, in a number of ways. We augment the particle physics model parameters with a series of nuisance parameters characterising the DM halo distribution, the most important SM masses and couplings, and the nuclear matrix elements relevant for the calculation of direct search yields. These are included in the scan as free parameters, and are constrained by a series of likelihoods derived from the best current knowledge of each observable (and in some cases, their correlations). Compared to the constraints used in Refs.~\cite{Cline13b, Beniwal}, we add improved direct detection likelihoods \cite{darkbit} from LUX \cite{LUXrun2}, PandaX \cite{PandaX2016}, SuperCDMS \cite{SuperCDMS} and XENON100 \cite{XENON2013}, as well as IceCube limits on DM annihilation to neutrinos in the core of the Sun \cite{IC79_SUSY,IC79}.  We also test some benchmark models obtained in our scan for stability of the electroweak vacuum.  Given the recent preference for astrophysical explanations of the \textit{Fermi}-LAT Galactic centre excess~\cite{Boyarsky:2010dr,Petrovic:2014uda,Carlson:2014cwa,Lee:2015fea,Bartels:2015aea,Cholis:2015dea,Clark16b}, we do not add this to the scan as a positive measurement of DM properties, unlike in Ref.~\cite{Cuoco:2016jqt}. We explore the extended parameter space in more detail than has previously been attempted, using four different scanning algorithms, and more stringent convergence criteria than previous studies.  The secondary purpose of this paper is to provide an example global statistical analysis using the Global and Modular Beyond-Standard Model Inference Tool (\gambit) \cite{gambit}, for a DM model where extensive comparison literature exists.

In Sec.~\ref{sec:phys}, we describe the Lagrangian and parameters of the scalar singlet model, discuss our astrophysical assumptions, and define the nuisance parameters that we include in our global fit.
Sec.~\ref{sec:scan} gives details of our scan, including the likelihood terms that we include for each constraint, the sampling algorithms we employ, and their settings. We present the latest status of the singlet model in Sec.~\ref{sec:results}, before concluding in Sec.~\ref{sec:conc}.

All input files, samples and best-fit benchmarks produced for this paper are publicly accessible from \textsf{Zenodo} \cite{the_gambit_collaboration_2017_801511}.

\section{Physics framework}
\label{sec:phys}

\subsection{Model definition}
\label{sec:model}

The renormalisable terms involving a new real singlet scalar $S$, permitted by the $\mathbb{Z}_2$, gauge and Lorentz symmetries, are
\begin{equation}
\mathcal{L} = \frac12 \mu_{\sss S}^2 S^2 + \frac12\lhs S^2|H|^2 + \frac14\ls S^4 + \frac12\partial_\mu S \partial^\mu S.
\end{equation}
From left to right, these are: the bare $S$ mass, the Higgs-portal coupling, the $S$ quartic self-coupling, and the $S$
kinetic term. Because $S$ never obtains a VEV, the model has only three free parameters: $\mu_{\sss S}^2$, $\lhs$ and $\ls$. Following electroweak symmetry breaking, the portal term induces $h^2S^2$, $v_0hS^2$ and $v_0^2S^2$ terms, where $h$ is the physical Higgs boson and $v_0 = 246$\,GeV is the VEV of the Higgs field.  The additional $S^2$ term leads to a tree-level singlet mass
\begin{equation}
\label{m_S_tree}
\ms = \sqrt{\mu_{\sss S}^2 + \frac12{\lhs v_0^2}}.
\end{equation}

Dark matter phenomenology is driven predominantly by $\ms$ and $\lhs$, with viable solutions known to exist \cite{Cline13b,Beniwal} in a number of regions: \begin{enumerate}
\item the resonance region around $\ms\sim\mh/2$, where couplings are very small ($\lhs<10^{-2}$) but the singlet can nevertheless constitute all of the observed DM,
\item the resonant ``neck'' region at $\ms=\mh/2$, with large couplings but an extremely small relic $S$ density, and
\item a high-mass region with order unity couplings.
\end{enumerate}

The parameter $\ls$ remains relevant when considering DM self-interactions (e.g.\ \cite{Tenkanen16}), and the stability of the electroweak vacuum. In the SM, the measured values of the Higgs and top quark masses indicate that the electroweak vacuum is not absolutely stable, but rather meta-stable \cite{Degrassi2012a}.  This means that although the present vacuum is not the global minimum of the scalar potential, its expected lifetime exceeds the age of the Universe.  Although this is not inconsistent with the existence of the current vacuum, one appealing feature of scalar extensions of the SM is that the expected lifetime can be extended significantly, or the stability problem solved entirely, by making the current vacuum the global minimum.

The stability of the electroweak vacuum has been a consideration in many studies of scalar singlet extensions to the SM \cite{Lerner09,Gonderinger2010,Drozd2011,Chen2012,Pruna13,Belanger2013a,Khan2014,Alanne2014,Han2015a,Kanemura2015,Robens15,Robens16}, typically appearing along with constraints from perturbativity, direct detection experiments and the relic abundance of DM.  As such, vacuum stability can be an interesting aspect to study of the scalar singlet model (and indeed, of any UV-complete model).  In this paper however, we primarily treat the scalar singlet DM model as a low-energy effective theory, and do not consider $\ls$ as a relevant parameter.  In a future fit, we plan to explore renormalisation of the scalar singlet model over the full range of scales, from electroweak to Planck, including full calculations of perturbativity and the lifetime of the electroweak vacuum.  Here, for the sake of interest we simply check the stability of the electroweak vacuum for a few of our highest-likelihood parameter points.

\subsection{Relic density and Higgs invisible width}
\label{sec:phys_rd}

In order to calculate the relic density of $S$, we need to solve the Boltzmann equation \cite{Gondolo:1990dk}
\begin{equation}
\label{eq:boltz}
\frac{dn_{\sss S}}{dt}+3Hn_{\sss S}=-\langle\sigma v_\mathrm{rel}\rangle\left(n_{\sss S}^2-n_{{\sss S},\mathrm{eq}}^2  \right)\,,
\end{equation}
where $n_{\sss S}$ is the DM number density, $n_{{\sss S},\mathrm{eq}}$ is the number density if the DM population were in chemical equilibrium with the rest of the Universe, $H$ is the Hubble rate, and
$\langle\sigma v_\mathrm{rel}\rangle$ is the thermally averaged self-annihilation cross-section times the relative velocity of the annihilating DM particles (technically the M{\o}ller velocity). The non-averaged cross-section $\sigma v$ depends on the centre-of-mass energy of the annihilation $\sqrt{s}$, and the thermal average depends on temperature $T$.  The average is given by
\begin{align}
\langle\sigma v_\mathrm{rel}\rangle=\int_{4\ms^2}^{\infty}ds\,\frac{s\sqrt{s-4\ms^2}K_1(\sqrt{s}/T)\sigma v_\mathrm{cms}} { 16 T \ms^4K_2^2(\ms/T)} \label{thermal_av}\,,
\end{align}
where for convenience we have expressed the result in terms of the relative velocity of the annihilating $S$ particles in the centre-of-mass frame, $v_{\rm cms}=2\sqrt{1-4\ms^2/s}$. For the case of the scalar singlet model, the non-averaged cross-section for annihilation into all final states except $hh$ is \cite{Cline13b}
\begin{align}
\sigma v_\mathrm{cms}=\frac{2\lhs^2v_0^2}{\sqrt{s}}
\frac{\Gamma_h(\sqrt{s}) }{ (s-\mh^2)^2+\mh^2\Gamma_h^2(\mh)}\,.
\label{eqn:sv}
\end{align}
For $\ms>\mh$, this expression needs to be supplemented with the partial annihilation cross-section into $hh$, given in Eq.\ A4 of Ref.\ \cite{Cline13b}.

We use the SM Higgs boson width $\Gamma_h(\sqrt{s})$ as a function of the invariant mass of the resonance $m_{h^*}=\sqrt{s}$, as implemented in \decaybit \cite{SDPBit}.\footnote{This comes from interpolating the results contained in the tables of Ref.\ \cite{YellowBook13}, and does not (yet) include theoretical uncertainties or the ability to recompute the width for different values of relevant nuisance parameters, such as $\alpha_s$ or quark masses.  Although included in \darkbit, we checked that it makes no difference to our results for $\ms<\mh/2$ whether or not we modify the width in the denominator of Eq.~\ref{eqn:sv} corresponding to the propagator of the internal Higgs, to take into account the decay channel $h\to SS$.}  At tree level, the decay width of Higgs bosons to such invisible final states is
\begin{equation}
  \Gamma_{h\rightarrow{\sss SS}} = \frac{\lhs^2 v_0^2}{32\pi \mh}\left(1 -4 \ms^2/\mh^2\right)^{1/2}\,.
  \label{Gamma_SS}
\end{equation}

This is the standard method for calculating the relic density.  It assumes that kinetic decoupling of DM from other species occurs well after chemical freeze-out.  If this is not the case, one must solve a coupled system of differential equations rather than the single Boltzmann equation (Eq.\ \ref{eq:boltz}) \cite{vandenAarssen:2012ag}. For the scalar singlet model, the standard approach is very accurate except at and below the Higgs resonance, where $\ms\sim \mh/2$.  Here, the impact of a more accurate treatment on the relic density can be up to one order of magnitude in the range
$53\,{\rm GeV}\lesssim m_\chi\lesssim63$\,GeV \cite{Binderinprep}, as $\sigma v_\mathrm{rel}$ is resonantly enhanced, so even small values of $\lhs$, where DM undergoes early kinetic decoupling, can avoid thermal overproduction.  Although this effect should arguably be included for the sake of completeness in future fits, it
has little impact on our final results because it only affects a relatively small mass range.

\subsection{Direct detection}

The predicted number of events in a direct detection experiment is
\begin{equation} \label{eqn:signal}
  N_\mathrm{p} = MT \int_0^{\infty} \phi(E) \frac{dR}{dE}(E) \, dE,
\end{equation}
with $M$ the detector mass, $T$ the exposure time, and $\phi(E)$ the detector response function.  The latter encodes the fraction of recoil events of some energy $E$ that are expected to be detected, within some analysis region.

The differential recoil rate $\frac{dR}{dE}$ depends on the nuclear scattering cross-section.  The scalar singlet model has no spin-dependent interactions with nuclei.  The spin-independent WIMP-nucleon cross-section is
\begin{equation}
\sigma_{SI}=\frac{m_N^4}{4\pi (\ms + m_N)^2}\frac{\lhs^2f_N^2}{\mh^4} ,
\end{equation}
where $m_N$ is the nucleon mass, and $f_N$ is the effective Higgs-nucleon coupling
\begin{equation}
f_N=\frac{2}{9}+\frac{7}{9}\sum_{q=u,d,s}f^{(N)}_{Tq}.
\end{equation}
The three light quark nuclear matrix elements $f^{(N)}_{Tq}$ can be calculated from the
nuclear matrix elements that describe the quark content of the proton and neutron,
\begin{align}
\sigma_l &\equiv m_l \langle N | \bar{u}u + \bar{d}d | N \rangle\,, \\
\sigma_s &\equiv m_s \langle N | \bar{s}s | N \rangle \, ,
\end{align}
where $m_l \equiv (1/2) (m_u + m_d)$, and $N \in \{p,n\}$.  See Ref.~\cite{Cline13b} for details.

Halo uncertainties can have a significant impact on the interpretation of direct searches for DM \cite{Benito:2016kyp}. For the DM halo in the Milky Way, we assume a generalised NFW profile, with a local Maxwell-Boltzmann speed distribution truncated at the local Galactic escape velocity.  The only parameter of the density profile that we retain as a nuisance parameter is $\rho_0$, the local DM density, although \GB makes it straightforward to also include uncertainties arising from the DM velocity profile in future fits.  For the scans of this paper, we assume a most probable speed $v_0=235$\,km\,s$^{-1}$ \cite{Reid:2009nj, Bovy:2009dr}, and an escape velocity of  $v_\mathrm{esc}=550$\,km\,s$^{-1}$ \cite{Smith:2006ym}. See Ref.~\cite{darkbit} for details.

\subsection{Indirect detection}

The flux of gamma rays from DM annihilation factorises into a part $\Phi$ that only
depends on the particle physics properties and a part $J$ that depends only on
the astrophysical distribution of DM.

For the gamma-ray flux in energy bin $i$ with width $\Delta E_i \equiv E_{\text{max},i} - E_{\text{min},i}$, the particle physics factor is
\begin{equation}
  \Phi_i = \sum_j\frac{\langle\sigma v\rangle_{0,j}}{8\pi \ms^2}\int_{E_\text{\text{min},i}}^{E_{\text{max},i}} dE \,
  \frac{dN_{\gamma,j}}{dE} \;,
  \label{eqn:gamLikePhi}
\end{equation}
where $dN_{\gamma,j}/dE$ is the differential gamma-ray multiplicity for single annihilations into final state $j$, and $\langle\sigma v\rangle_{0,j}\equiv \sigma v_j|_{v\to0} \equiv \sigma v_j|_{s\to4\ms^2}$ is the zero-velocity
limit of the partial annihilation cross-section into final state $j$.  This is the final-state-specific equivalent of Eq.~\ref{eqn:sv}.  We compute the partial annihilation cross-sections for the singlet model using the expressions of Appendix A of Ref.~\cite{Cline13b}, as implemented in \darkbit \cite{darkbit}. We obtain the predicted spectra ${dN_{\gamma}}/{dE}$ for each model point by using a Monte-Carlo showering simulation,
detailed in Ref.~\cite{darkbit}.

The astrophysics factor for a given target $k$ is
\begin{equation}
  \label{eq:astroJ}
  J_k = \int_{\Delta \Omega_k} d\Omega\int_\text{l.o.s.} ds \, \rho_{\sss S}^2\;.
\end{equation}
Here, $\Delta \Omega_k$ denotes the solid angle over which the signal is integrated, l.o.s. indicates that the line element $ds$ runs along the line of sight to the target object, and
$\rho_{\sss S}$ is the DM mass density within it.

Neutrino telescopes also place bounds on DM models by searching for high-energy neutrinos from DM annihilation.  The most likely signal in this respect comes from DM gravitationally captured by the Sun and concentrated to its core, where it would annihilate \cite{Bergstrom98b, IC79}.  This channel predominantly tests the mass and couplings of DM to nuclei rather than the annihilation cross-section, as the nuclear scattering leading to capture is the rate-limiting step for most models.  Because the singlet model has no spin-dependent couplings to nuclei, neutrino telescope searches for annihilation in the Sun provide constraints only on the spin-independent scattering cross-section.

Owing to the uncertainties associated with cosmic ray propagation, we do not consider constraints from charged cosmic rays (primarily anti-protons and positrons).  Radio signals coming from synchrotron emission by DM annihilation products generated in strong magnetic fields are not included in our analysis, as the associated field strengths are highly uncertain.  Nor are CMB limits on DM annihilation, as \textit{Fermi} dwarf limits are stronger at all masses of interest for this model.  Finally, we do not consider limits implied by gamma-ray observations of the Galactic centre, whether by \textit{Fermi} or ground-based gamma-ray telescopes, owing to the uncertainties involved in modelling the DM profile and astrophysical gamma-ray emission of the central Milky Way.

\section{Scan details}
\label{sec:scan}

\begin{table}
\begin{center}
\caption{\label{tab:param} Scalar singlet model parameters varied in our fits, along with their associated ranges and prior types.}
\begin{tabular}{l c c c}
\hline
Parameter & Minimum & Maximum & Prior    \\
\hline
$\lhs$                   & $10^{-4}$ & 10      & log  \\
$\ms$ (full-range scan)  & 45\,GeV   & 10\,TeV & log  \\
$\ms$ (low-mass scan)    & 45\,GeV   & 70\,GeV & flat \\
\hline
\end{tabular}
\end{center}
\end{table}

\begin{table}[tp]
\centering
\caption{Names and ranges of Standard Model, halo and nuclear nuisance parameters that we vary simultaneously with scalar singlet parameters in our fits.  We assign a flat prior to all these parameters.} \label{tab:SMparams}
\begin{tabular}{l@{\hspace{-5mm}}c@{\,}r}
\hline
Parameter & & Value($\pm$Range) \\
\hline
Local DM density & \phantom{$^{\MSBar}$}$\rho_0$\phantom{$^{\MSBar}$} &  0.2--0.8\,GeV\,cm$^{-3}$\\
Nuclear matrix el. (strange)  & \phantom{$^{\MSBar}$}$\sigma_s$\phantom{$^{\MSBar}$} & $43(24)$\,MeV \\
Nuclear matrix el. (up + down) & \phantom{$^{\MSBar}$}$\sigma_l$\phantom{$^{\MSBar}$} & $58(27)$\,MeV \\
Strong coupling & $\alpha_s^{\MSBar}(m_Z)$      & $0.1185(18)$   \\
Electromagnetic coupling & $1/\alpha^{\MSBar}(m_Z)$        & $127.940(42)$       \\
Fermi coupling $\times$ $10^{5}$ & \phantom{$^{\MSBar}$}$G_{F,5}$\phantom{$^{\MSBar}$} & $1.1663787(18)$ \\
Higgs pole mass & \phantom{$^{\MSBar}$}$\mh$\phantom{$^{\MSBar}$}  & 124.1--127.3\,GeV \\
Top pole mass  & \phantom{$^{\MSBar}$}$m_t$\phantom{$^{\MSBar}$}  &  $173.34(2.28)$\,GeV\\
Bottom quark mass & $m_b^{\MSBar}(m_b)$    & $4.18(9)$\,GeV \\
Charm quark mass & $m_c^{\MSBar}(m_c)$ & $1.275(75)$\,GeV \\
Strange quark mass & $m_s^{\MSBar}(2\,\text{GeV})$  & $95(15)$\,MeV\\
Down quark mass & $m_d^{\MSBar}(2\,\text{GeV})$  &   $4.80(96)$\,MeV  \\
Up quark mass & $m_u^{\MSBar}(2\,\text{GeV})$     & $2.30(46)$\,MeV \\
\hline\end{tabular}
\end{table}

\subsection{Parameters and nuisances}

A summary of the parameter ranges that we scan over for this paper is given in Tables~\ref{tab:param} and \ref{tab:SMparams}.

Table \ref{tab:param} gives the singlet model parameters, along with the scanning priors that we use. We carry out two main types of scan: one over the full range of masses from 45\,GeV to 10\,TeV, intended to sample the entire parameter space, and another centred on lower masses at and below the Higgs resonance $\ms\sim \mh/2$, in order to obtain a more detailed picture of the resonance region.

In addition to the effect of the singlet parameters, we also consider the effects of varying a number of SM, astrophysical and nuclear parameters within their allowed experimental uncertainties. Table \ref{tab:SMparams} gives the full ranges of the 13 nuisance parameters that we vary in our scans, along with their central values.  We assign flat priors to all nuisance parameters in Table~\ref{tab:SMparams}, as they are all sufficiently well constrained that their priors are effectively irrelevant.

We allow for $\pm3\sigma$ excursions from the best estimates of the nuclear couplings.  For the local DM density, we scan an asymmetric range about the central value, reflecting the log-normal likelihood that we apply to this parameter (Sec.\ \ref{like_nuisance}).  Detailed references for the central values and uncertainties of these parameters can be found in Ref.\ \cite{darkbit}.

The central values of the up and down quark masses come from the 2014 edition of the PDG review \cite{PDB}; we allow these parameters to vary by $\pm20\%$ in our fits, so as to encompass the approximate $3\sigma$ range of correlated uncertainties associated with the mass ratio likelihoods implemented in \precisionbit \cite{SDPBit}.  Given the large impact that the Higgs mass can have on the phenomenology of this model, we scan an extended range for this parameter, covering more than $\pm4\sigma$ around the central value quoted in the 2015 update to the PDG review \cite{PDG15} ($m_h=125.09\pm0.24$\,GeV; see Sec.\ \ref{like_nuisance}).  The central value and $\pm3\sigma$ scan range for the top quark pole mass come from Ref.~\cite{ATLAS:2014wva}, and for all other SM nuisance parameters from Ref.~\cite{PDB}.

We include the local DM density and nuclear matrix elements as nuisance parameters because of their impacts on direct detection and capture of singlet particles by the Sun.  The strong coupling, Higgs VEV (determined by $G_F$), Higgs mass and quark masses all enter into the cross-sections for annihilation and/or scattering of $S$ \cite{Cline13b}.  The electromagnetic coupling does not impact our fit beyond its own nuisance likelihood, but has a small effect on renormalisation of other parameters and therefore vacuum stability, which we investigate for a few benchmarks and will explore in detail in a follow-up paper.

\subsection{Scanning procedure}

Although 13 of the directions in the 15-dimensional parameter space are well constrained, efficiently sampling all 15 parameters simultaneously requires sophisticated scanning algorithms.  We explore this space primarily with two different scanning packages interfaced via \scannerbit: a differential evolution sampler \diver, and an ensemble Markov Chain Monte Carlo (MCMC) known as \twalk \cite{scannerbit}.  Both algorithms are the current state of the art when it comes to scaling with dimension \cite{scannerbit}, and thus are the natural choice for this study.

Both of these algorithms are particularly well suited for multimodal distributions, and each serves a purpose in this study.  \twalk allows efficient and accurate calculation of the Bayesian posterior distribution for the target model. The package can also be used for frequentist studies if the sampling density is amplified by a judicious choice of run parameters.  However, \twalk is far less efficient at sampling the profile likelihood in high-dimensional spaces than \diver \cite{scannerbit}.  Because we vary 15 parameters in total, we use \diver to produce high-quality profile likelihoods.  Having identified all likelihood modes, and therefore all possible locations that might meaningfully contribute to the posterior, we then use \twalk to produce posterior distributions, checking that it does not fail to locate any of the modes identified by \diver.

In addition to the ensemble MCMC and differential evolution scans, we also combine our results with those from a more traditional MCMC, \great, and the nested sampling algorithm \multinest.  These are also interfaced to \scannerbit \cite{scannerbit}.  Although it is not typically necessary to combine results from four different algorithms, here we demonstrate the power of the \gambit package, which allows us to use a range of scanning procedures on the same composite likelihood, in order to produce the most robust results possible.

As discussed in Sec.\ \ref{sec:model}, the singlet parameter space features a viable region at $\ms \approx \mh/2$.  In this region, the annihilation of singlet DM to SM particles via s-channel Higgs exchange is resonantly enhanced, and a lower portal coupling is required to achieve the observed relic density.  This region is not yet excluded by direct detection.  However, probing this region of the parameter space over a large-mass range is difficult, even when using a logarithmic prior on the mass.  To properly sample this region, we run a second scan with each sampler, using a flat prior over the range $\ms\in [45,70]$\,GeV.  We also carry out an additional specially focused low-mass scan with \diver in the ``neck'' region of the resonance, in order to obtain well-sampled contours in the most localised part of the allowed parameter space.  We do this by excluding all points outside the range $\ms\in [61.8,63.1]$\,GeV.

The convergence criteria, population size and chain details are controlled by various settings for each sampler.  The settings that we use in this paper are presented in Table \ref{table:scanner_params}.  We chose these settings after extensive testing \cite{scannerbit}, to give the most stringent convergence and best exploration possible with each scanner and region.  To a certain extent, some of these settings are overkill for the problem at hand, and the same physical inference could be achieved with less samples.  However, the scans that we present here took only 26\,000 core hours in total to compute, and the scan that dominates most of the contours (the full-range \diver scan) took just 3\,hr on 10 $\times$ 24-core nodes, i.e. around 700 core hr.  Compared to the time required to compute fits that include direct LHC simulations \cite{ColliderBit,CMSSM,MSSM}, the additional sampling we do here costs practically nothing -- and noticeably improves the resolution of our results.  We refer the reader to Ref.\ \cite{scannerbit} for further details of the scanners, their settings and underlying algorithms.

The profile likelihoods that we present in this paper are based on the combination of all samples from all scans, which contain $5.7\times10^7$ valid samples altogether.  In contrast, the posteriors that we show come exclusively from the full-range \twalk scan.

We compute and plot profile likelihoods and posteriors using \pippi \cite{pippi}, obtaining profile likelihoods by maximising the log-likelihood in parameter bins over all other parameters not shown in a given plot, and posteriors by integrating the posterior density over the parameters not shown in each plot.  We compute confidence regions and intervals by determining the appropriate iso-likelihood contour relative the best-fit likelihood for 1 or 2 degrees of freedom, corresponding to 1D and 2D plots, respectively.  We compute Bayesian credible regions and intervals as parameter ranges containing the relevant posterior mass according to the maximum posterior density requirement.  Further details can be found in Ref.\ \cite{pippi}.

\begin{table}[tp]
\caption{Parameters of each sampler for carrying out global fits of the scalar singlet model in this paper.}\label{table:scanner_params}
\centering
\begin{tabular}{l l l l}
\hline
Scanner & Parameter & Full range & Low mass\\
\hline
\MultiNest & \cpp{nlive} & 20,000 & 20,000 \\
  & \cpp{tol} & $10^{-4}$ & $10^{-5}$ \\
\hline
\diver & \fortran{NP} & 50,000 & 50,000 \\
  & \fortran{convthresh} & $10^{-4}$  & $10^{-5}$\\
  \hline
\twalk & \cpp{chain_number} &$ 512$ & $512$ \\
  & \cpp{sqrtR} $-$ 1 & $0.01$ & $0.01$ \\
    \hline
\great & \cpp{nTrialLists} &17 & 17\\
  & \cpp{nTrials} & 20,000 & 10,000\\
\hline\end{tabular}
\end{table}

\subsection{Relic density likelihood}

To determine the thermal $S$ relic density for each parameter combination, we solve the Boltzmann equation (Eq.\ \ref{eq:boltz}) numerically with \darkbit \cite{darkbit}, taking the partial annihilation rates for different final states from Eq.\ \ref{eqn:sv} supplemented at $\ms > \mh$ with the expression for $\langle\sigma v\rangle_{0,hh}$ from Appendix A of Ref.~\cite{Cline13b}. For $\ms < 150$\,GeV we use the SM Higgs partial widths contained in \decaybit (from Ref.\ \cite{YellowBook13}), whereas for $\ms > 150$\,GeV we revert to the tree-level expressions from Appendix A of Ref.~\cite{Cline13b}, to avoid the impact of large 1-loop corrections to the Higgs self-interaction.  We determine the effective invariant rate $W_{\rm eff}$ from the partial annihilation cross-sections, and pass it on to the numerical Boltzmann solver of \ds~\cite{darksusy} in order to obtain $\Omega_{\sss S} h^2$.

We implement the relic density likelihood as an upper limit only, permitting models where the thermal abundance makes $S$ a fraction of DM.  Comparing with the relic abundance measured by \textit{Planck} \cite{Planck15cosmo}  ($\Omega_\text{DM} h^2 = 0.1188\pm 0.0010$, at $1\sigma$), we compute a marginalised Gaussian upper limit likelihood as described in Sec 8.3.4 of Ref.\ \cite{gambit}.  Models that predict less than the measured relic density are assigned a likelihood contribution equal to that assigned to models that predict the observed value exactly.  Models predicting more than the measured relic density are penalised according to a Gaussian function centred on the observed value.  We adopt the \darkbit default value of 5\% for the theoretical uncertainty on the relic density prediction, adding it in quadrature to the experimental uncertainty on the observed value.  We note that this is a very conservative estimate of the theoretical uncertainty for this model, except in the resonance region (see Sec.\ \ref{sec:phys_rd}).

For models that underpopulate the observed relic density, we rescale all direct and indirect signals to account for the fraction of DM that is detectable using the properties of the $S$ boson.  This is internally consistent from the point of view of the model, and conservative in the sense that it suppresses direct and indirect signals in regions where the thermal abundance is less than the \textit{Planck} value.

\subsection{LHC Higgs likelihoods}

When $\ms<\mh/2$, the decay $h\rightarrow S S$ is kinematically allowed, with a partial width given by Eq.~\ref{Gamma_SS}. This is entirely invisible at hadron colliders.  Constraints can be placed on the scalar singlet model parameters from measurements of Higgs production and decay rates, and the implied limit on invisible decay channels of the Higgs. For the case of SM-like couplings, the 95\% confidence level upper limit on the Higgs invisible width from LHC and Tevatron data is presently at the level of 19\% \cite{Belanger:2013xza}. We use the \decaybit implementation of the complete invisible Higgs likelihood, based on an interpolation of Figure 8 of~\cite{Belanger:2013xza}.

\subsection{Direct detection likelihoods}
\label{sec:dd_like}

The dominant constraints on the scalar singlet model come from the LUX \cite{LUX2016,LUXrun2} and PandaX \cite{PandaX2016} experiments, with weaker limits also available from \darkbit based on SuperCDMS \cite{SuperCDMS} and XENON100~\cite{XENON2013}. We use the \darkbit interface to \ddcalc to evaluate a Poisson likelihood for observing $N_{\mathrm{o}}$ events in a given experiment, given a predicted number of signal events $N_{\mathrm{p}}$ (Eq.~\ref{eqn:signal}),
\begin{equation} \label{eqn:Poisson}
  \mathcal{L}(N_{\mathrm{p}}|N_{\mathrm{o}})
  = \frac{(b+N_{\mathrm{p}})^{N_{\mathrm{o}}} \, e^{-(b+N_{\mathrm{p}})}}{N_{\mathrm{o}}!}.
\end{equation}
Here $b$ is the expected number of background events in the analysis region.  We model detector efficiency and acceptance effects by interpolating between values in pre-calculated tables contained in \ddcalc.

\subsection{Indirect detection likelihoods}

The lack of evidence for anomalous gamma-ray emission from dwarf spheroidal galaxies in data collected by the \textit{Fermi}-LAT experiment allows stringent constraints to be placed on the DM annihilation cross-section~\cite{LATdwarfP8}. We use the \texttt{Pass 8} analysis of the 6-year dataset, with the composite likelihood
\begin{equation}
  \ln\mathcal{L}_\text{exp} = \sum_{k=1}^{N_\text{dSph}}\sum_{i=1}^{N_\text{ebin}}
  \ln\mathcal{L}_{ki}(\Phi_i \cdot J_k)\;,
  \label{eqn:FermiTabLike}
\end{equation}
where $N_\text{dSph}$ and $N_\text{ebin}$ are the number of considered
dSphs and the number of energy bins, respectively.  The partial likelihoods $\mathcal{L}_{ki}$ are functions of the signal flux,
and hence of the quantities $\Phi_i$ and $J_k$ defined in Eqs.~\ref{eqn:gamLikePhi} and \ref{eq:astroJ}, respectively.

The main results of Ref.~\cite{LATdwarfP8} were obtained by profiling over the $J_k$ as nuisance parameters, yielding a combined profile likelihood of
\begin{equation}
  \ln \mathcal{L}_\text{dwarfs}^\text{prof.}(\Phi_i) = \max_{J_1\dots
  J_k}\left(\ln\mathcal{L}_\text{exp} + \ln\mathcal{L}_J\right),
\end{equation}
where
\begin{equation}
  \ln\mathcal{L}_J = \sum_{k=1}^{N_\text{dSph}} \ln\mathcal{N}(\log_{10} J_k |
    \log_{10} \hat J_k ,
  \sigma_k).
\end{equation}
Here the use of a log-normal distribution to describe the uncertainty on $J_k$ is a good approximation. Tabulated binned likelihoods have been provided by the \textit{Fermi}-LAT experiment, and implemented in \darkbit via the \gamlike package.\footnote{\url{https://www-glast.stanford.edu/pub\_data/1048/}}

The strongest neutrino indirect detection constraints on DM-nucleon scattering currently come from the IceCube search for annihilation in the Sun \cite{IC79,IC86}.  We access the 79-string results via the \darkbit interface to the \nulike package \cite{IC79_SUSY,IC22Methods}, which constructs a fully unbinned likelihood using event-level energy and angular information available in the published 79-string IceCube dataset, marginalised over detector systematics.  We obtain predicted neutrino spectra at the Earth using \textsf{WimpSim} \cite{Blennow08} yield tables contained in \darksusy \cite{darksusy}.  Although IceCube limits on spin-independent scattering are not competitive with those from LUX or PandaX, for many points in the singlet parameter space they provide constraints stronger than those from SuperCDMS, and almost as strong as XENON100.

Note that the methods that we use for marginalising or profiling out additional systematic uncertainties in neutrino and $\gamma$-ray likelihoods are only applicable because the systematics are uncorrelated; the same cannot be done with a common systematic that impacts many experiments, such as the local density of DM (which affects every direct detection experiment).

The dwarf likelihood gives an identical result to what we would obtain if we were to include each of the $J$ factors as nuisance parameters in our own fit, and profile over them.  The same is true of the IceCube detector systematics treated by \textsf{nulike}, although in that case the equivalent result would be the Bayesian one, where the corresponding nuisance parameter was marginalised over.  Ideally, one would include all such nuisance parameters in the same fit, and then be free to choose at the end of a scan to profile over them all to produce profile likelihoods, or marginalise over them all to produce posterior probability densities.  In practice however, the gain in accuracy achieved by doing so is generally minimal, whereas the speed gain from the `inline' treatment is substantial.

\subsection{Nuisance likelihoods}
\label{like_nuisance}

Following Ref.~\cite{darkbit}, we take the likelihood terms for the hadronic matrix elements $\sigma_s$ and $\sigma_l$ to be Gaussian, with central values and $1\sigma$ uncertainties of $43\pm8$ and $58\pm9$, respectively.

The canonical value of the local DM density $\rho_0$ is $\bar{\rho_0}=0.4$ GeV/cm$^3$ (e.g.\ \cite{Catena:2009mf}), but this depends on assumptions such as spherical symmetry in the halo. We remain relatively agnostic with respect to this assumption by choosing a log-normal distribution for the likelihood of $\rho_0$, and assuming an uncertainty of $\sigma_{\rho_0} = 0.15$\,GeV\,cm$^{-3}$, such that
\begin{equation}
\mathcal{L}_{\rho_0} =  \frac{1}{\sqrt{2\pi} \sigma'_{\rho_0} \rho_0} \exp \left(- \frac{\ln(\rho_0 / \bar\rho_0)^2}{2 {\sigma^{\prime 2}_{\rho_0}}} \right) ,
\end{equation}
where $\sigma'_{\rho_0} = \ln(1 + \sigma_{\rho_0}/\rho_0)$. More details can be found in Ref.~\cite{darkbit}.

We use the \precisionbit implementation of SM nuisance parameter likelihoods. For the \MSbar light quark $(u,d,s)$ masses at $\mu=2$~GeV, we use a single joint Gaussian likelihood function, combining likelihoods on $m_u/m_d$, $m_s/(m_u+m_d)$, and $m_s$. We take the experimental measurements of these quantities and their uncertainties from the PDG~\cite{PDB}.  We use Gaussian likelihoods for $G_F$, based on the measured value $G_{F} = (1.1663787 \pm 0.0000006) \times 10^{-5}$ GeV$^{-2}$, $\alpha_{\rm EM}$, based on the observed $\alpha_{\mathrm{EM}}(m_{Z})^{-1} = 127.940 \pm 0.014$ (\MSbar scheme)~\cite{PDB}, and $\alpha_{s}$, using the value $\alpha_{s}(m_{Z}) = 0.1185 \pm 0.0005$ (\MSbar scheme), as obtained from lattice QCD.~\cite{PDB}. We use the quoted uncertainties as $1\,\sigma$ confidence intervals, and apply no additional theoretical uncertainty.  We also apply a simple Gaussian likelihood with no theoretical uncertainty to the Higgs mass, based on the 2015 PDG result of $\mh = 125.09 \pm 0.24$ GeV \cite{PDG15}.

\begin{figure*}[t]
\centering
\includegraphics[height=0.85\columnwidth]{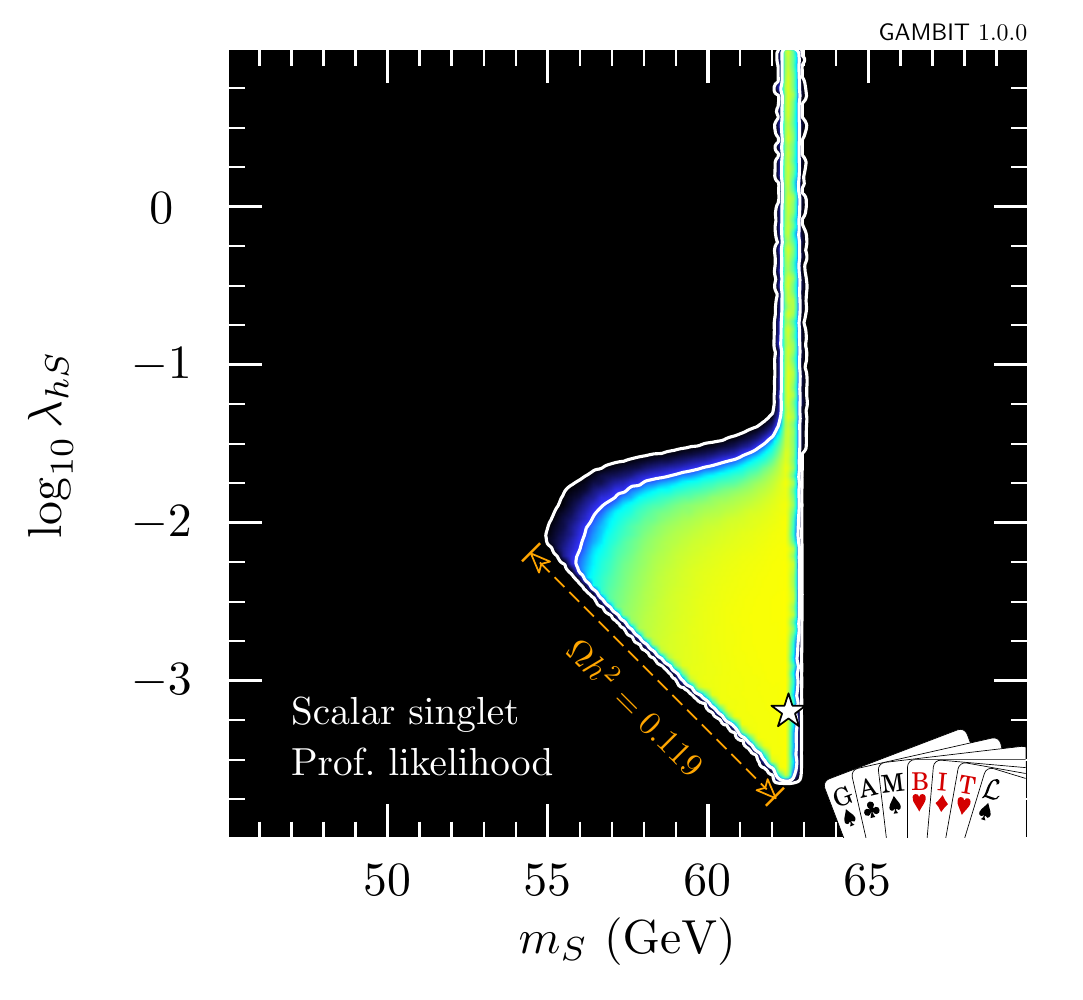}\hspace{0.05\columnwidth}\includegraphics[height=0.85\columnwidth]{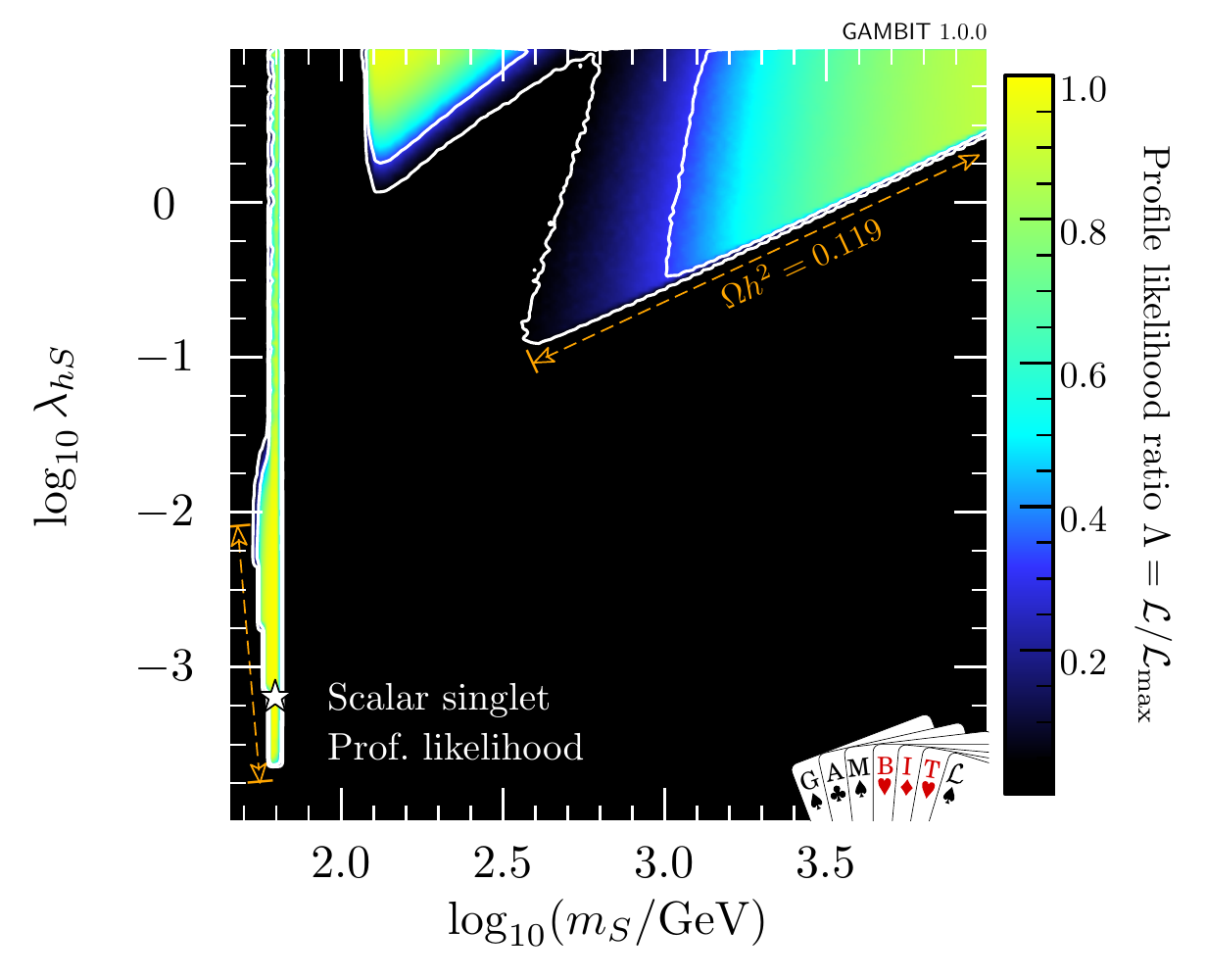}
\caption{Profile likelihoods for the scalar singlet model, in the plane of the singlet parameters $\lhs$ and $\ms$.  Contour lines mark out the $1\sigma$ and $2\sigma$ confidence regions.  The left panel shows the resonance region at low singlet mass, whereas the right panel shows the full parameter range scanned.  The best-fit (maximum likelihood) point is indicated with a white star, and edges of the allowed regions corresponding to solutions where $S$ constitutes 100\% of dark matter are indicated in orange.}
\label{fig::Ms_lhs}
\end{figure*}

\begin{figure*}[t]
\centering
\includegraphics[width=0.7\columnwidth]{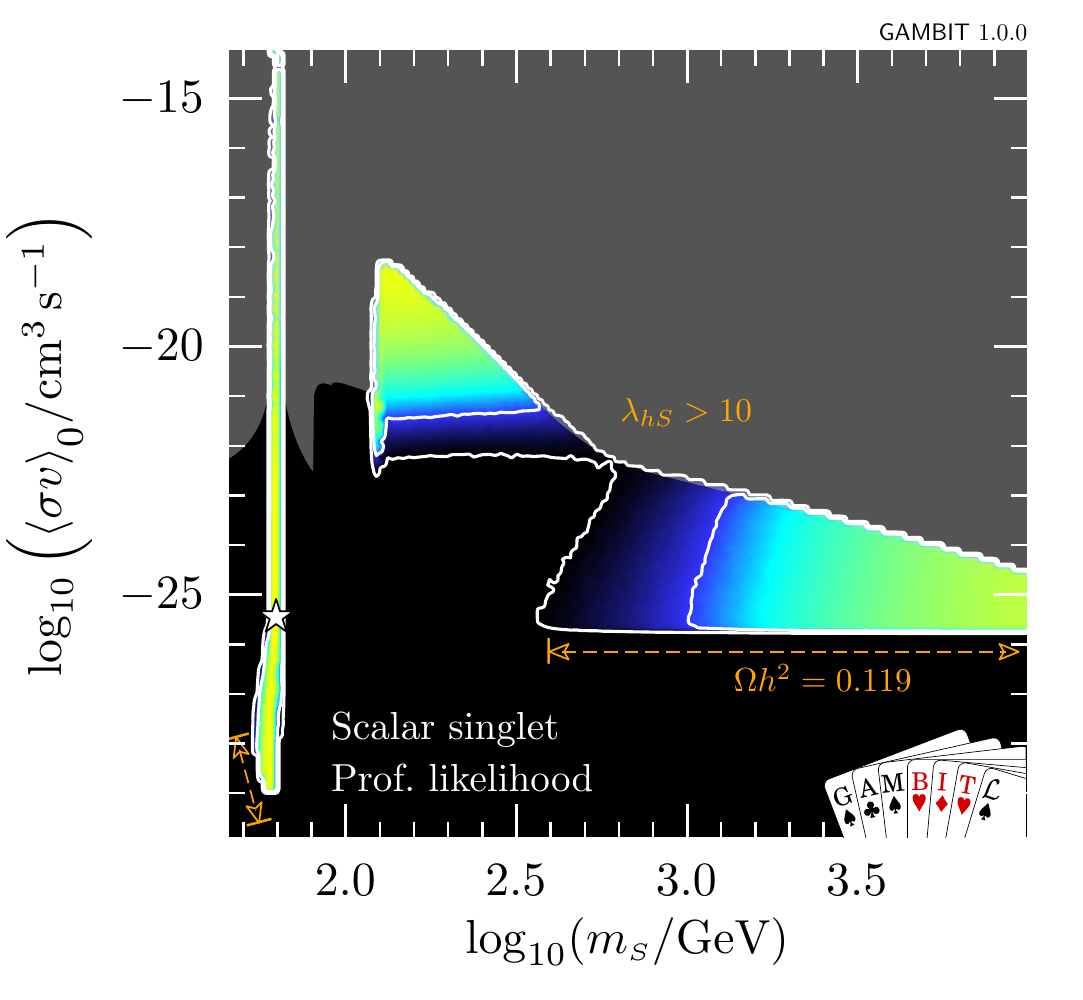}\includegraphics[width=0.7\columnwidth]{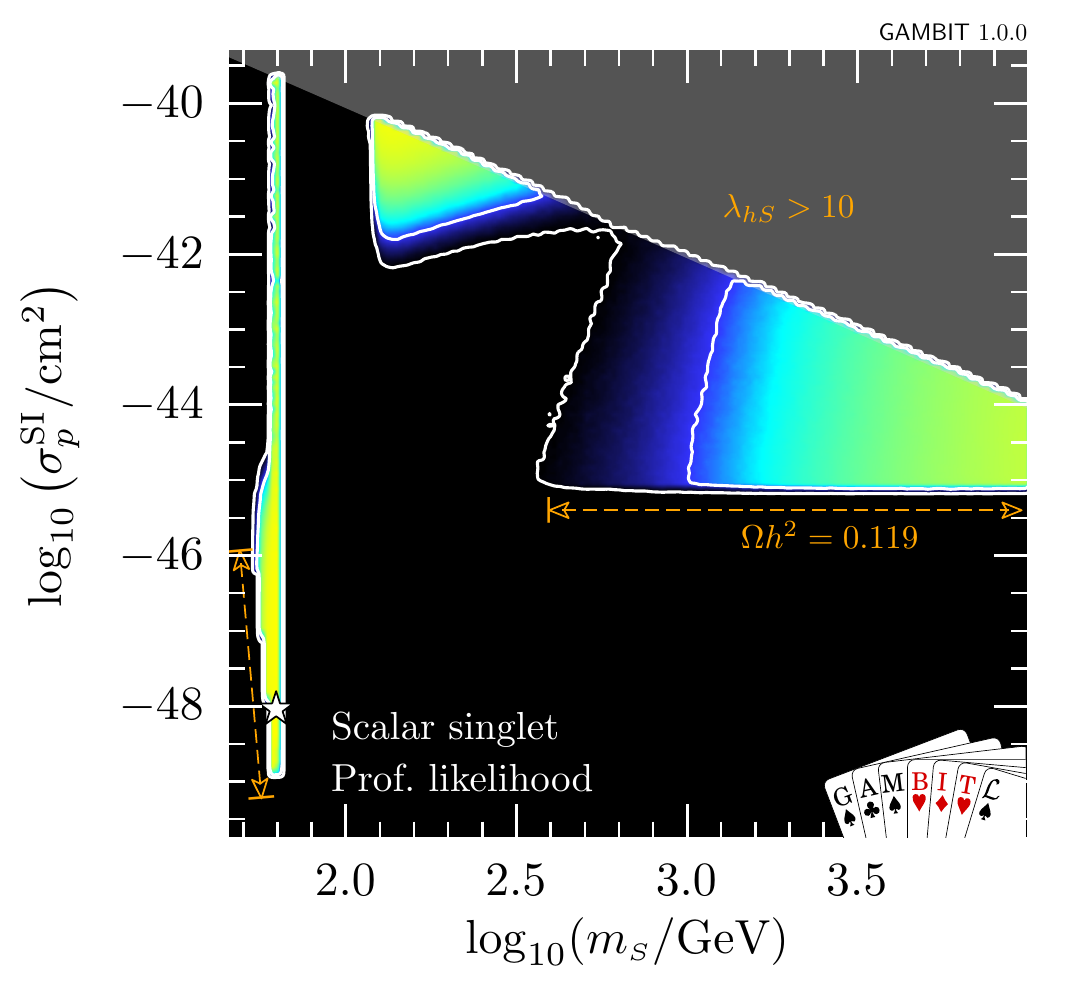}\includegraphics[width=0.8\columnwidth]{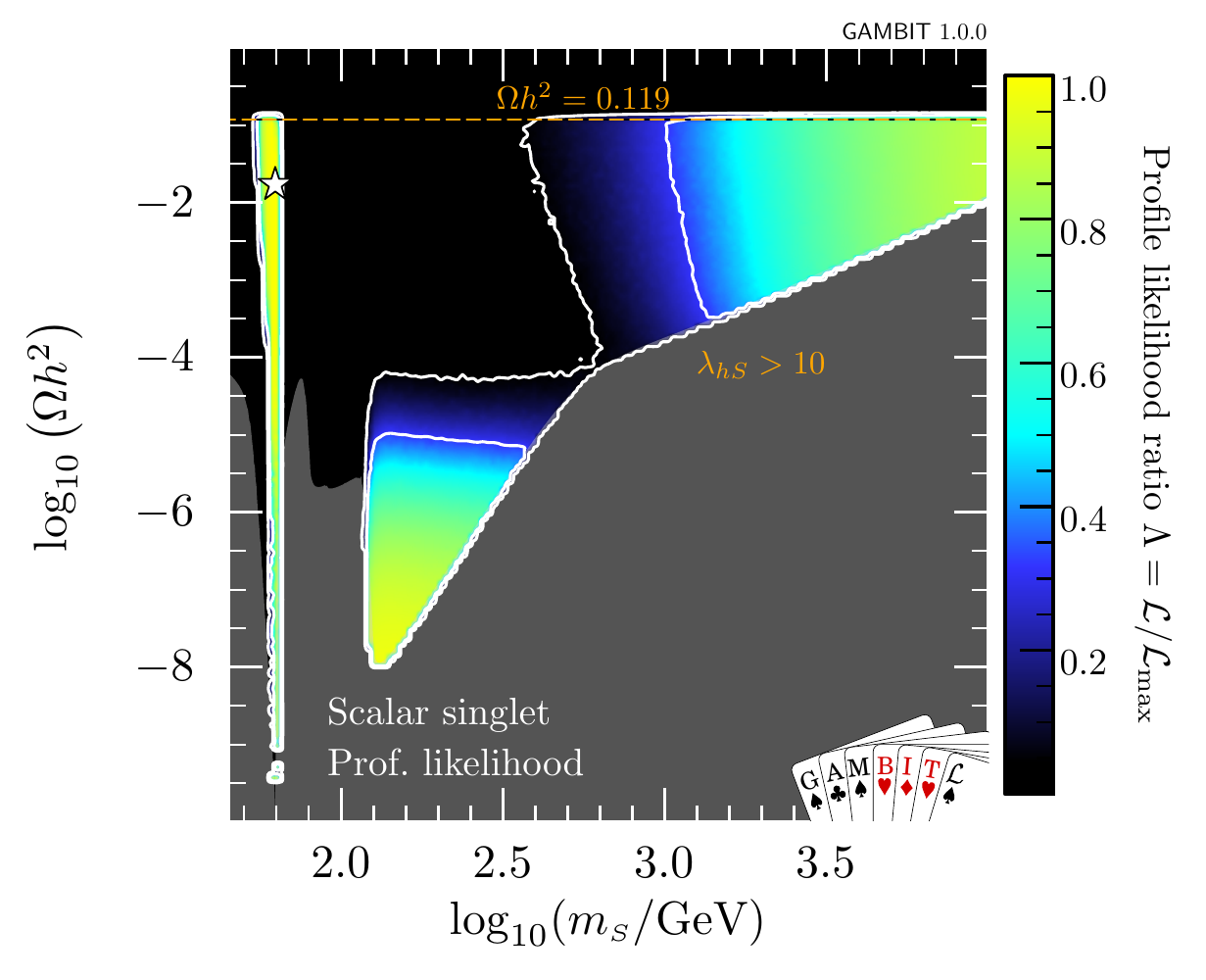}
\caption{Profile likelihoods for the scalar singlet model, in various planes of observable quantities against the singlet mass.  Contour lines mark out the $1\sigma$ and $2\sigma$ confidence regions.  Greyed regions indicate values of observables that are inaccessible to our scans, as they correspond to non-perturbative couplings $\lhs>10$, which lie outside the region of our scan.  Note that the exact boundary of this region moves with the values of the nuisance parameters, but we have simply plotted this for fixed central values of the nuisances, as a guide.  The best-fit (maximum likelihood) point is indicated with a white star, and edges of the allowed regions corresponding to solutions where $S$ constitutes 100\% of dark matter are indicated in orange.
\textit{Left:}  late-time thermal average of the cross-section times relative velocity;
\textit{Centre:}  spin-independent WIMP-nucleon cross-section;
\textit{Right:} relic density.}
\label{fig::Ms_oh2}
\end{figure*}

\begin{figure*}[t]
\centering
\includegraphics[height=0.85\columnwidth]{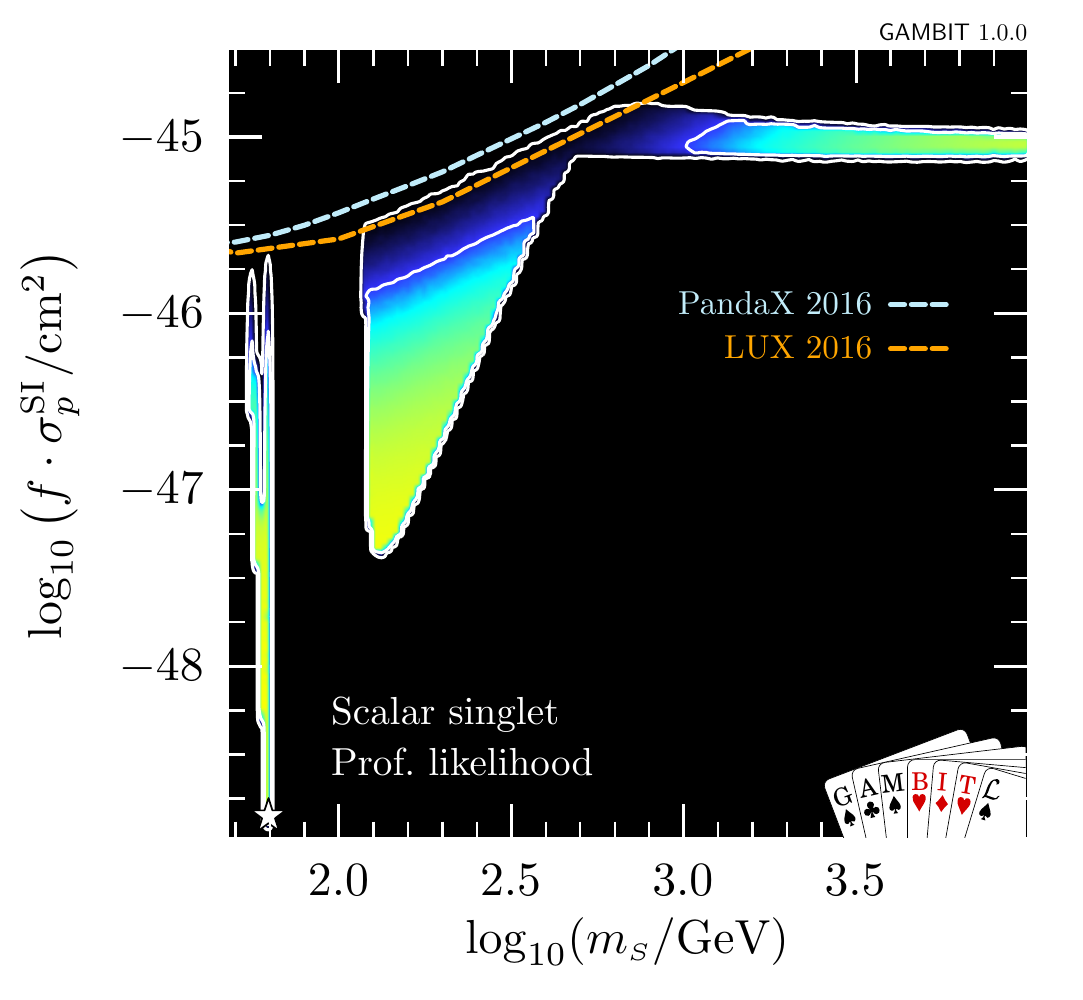}\hspace{0.05\columnwidth}\includegraphics[height=0.85\columnwidth]{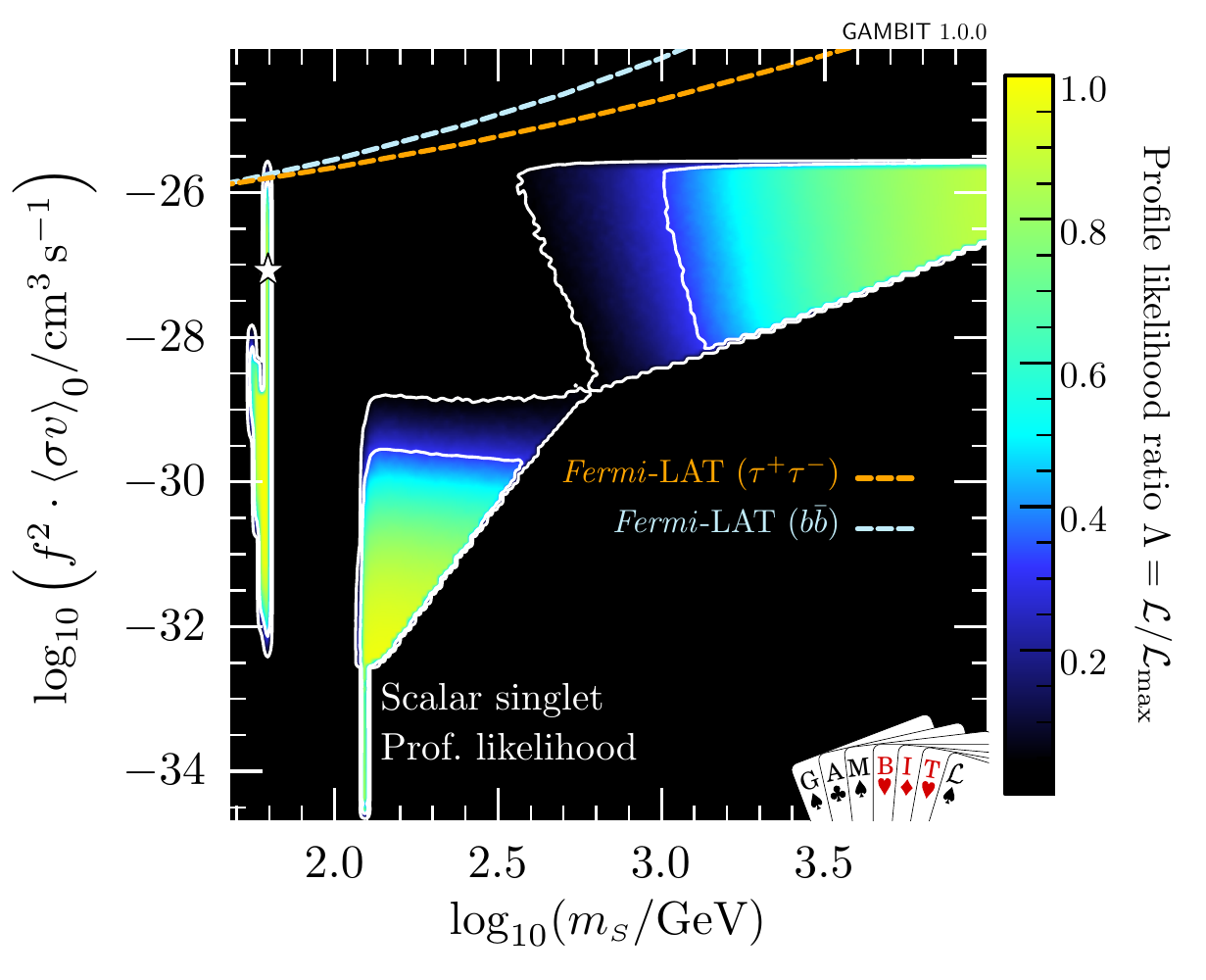}
\caption{Profile likelihoods of nuclear scattering (\textit{left}) and annihilation (\textit{right}) cross-sections for the scalar singlet model, scaled for the singlet relic abundance and plotted as a function of the singlet mass.  Here we rescale the nuclear and annihilation scattering cross-sections by $f\equiv\Omega_{\sss S} / \Omega_\text{DM}$ and $f^2$, in line with the linear and quadratic dependence, respectively, of scattering and annihilation rates on the dark matter density.  Contour lines mark out the $1\sigma$ and $2\sigma$ confidence regions.  The best-fit (maximum likelihood) point is indicated with a white star.}
\label{fig::Ms_oh2_scaled}
\end{figure*}

\begin{figure*}[t]
\includegraphics[width=0.212\textwidth]{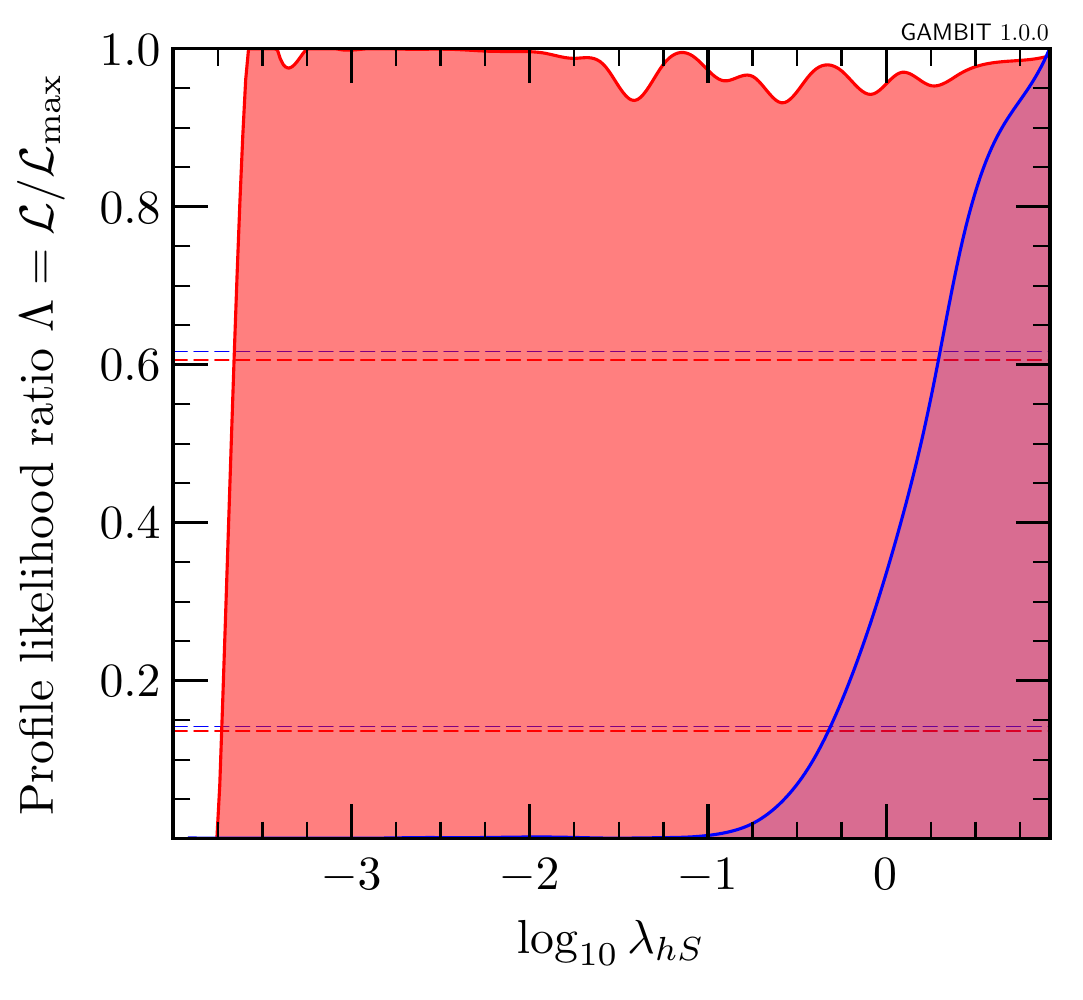}\hspace{-3mm}%
\includegraphics[width=0.212\textwidth]{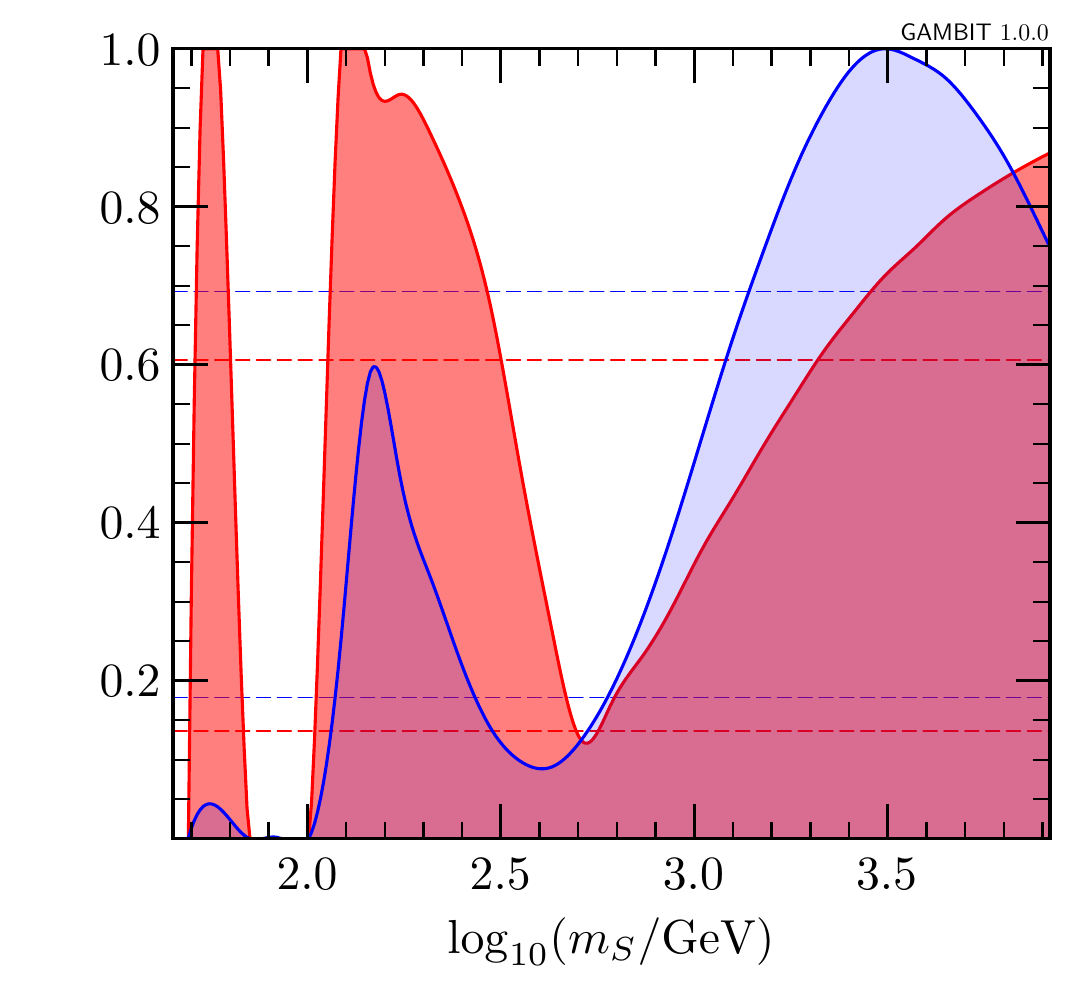}\hspace{-3mm}%
\includegraphics[width=0.212\textwidth]{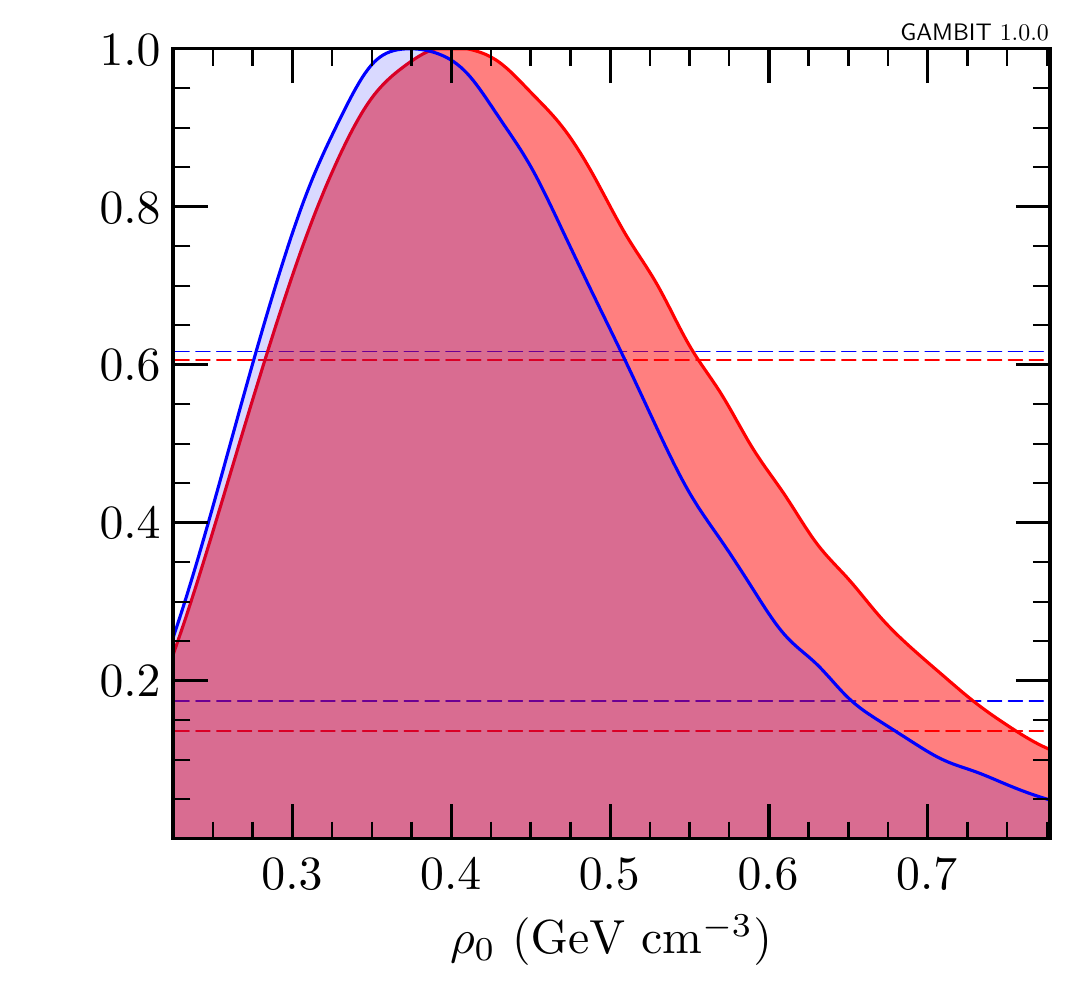}\hspace{-3mm}%
\includegraphics[width=0.212\textwidth]{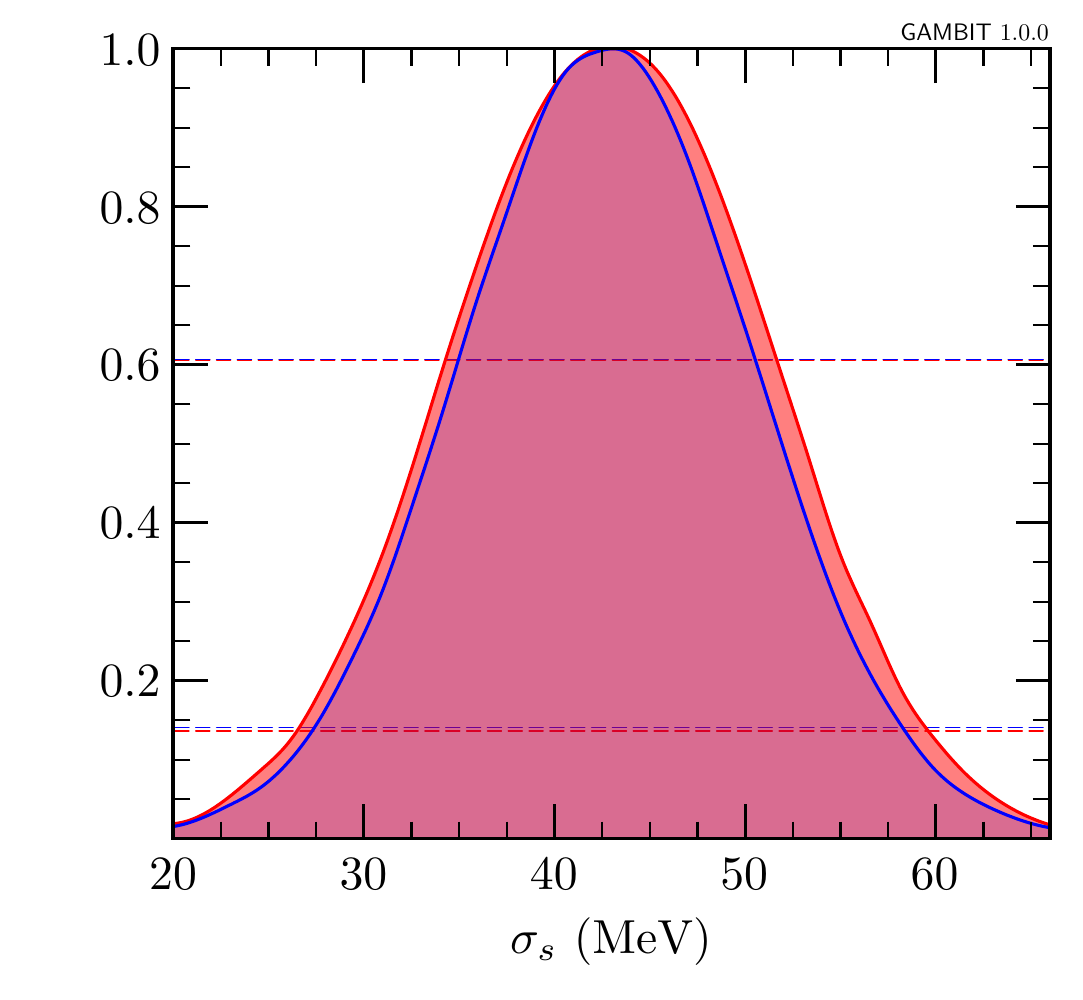}\hspace{-3mm}%
\includegraphics[width=0.212\textwidth]{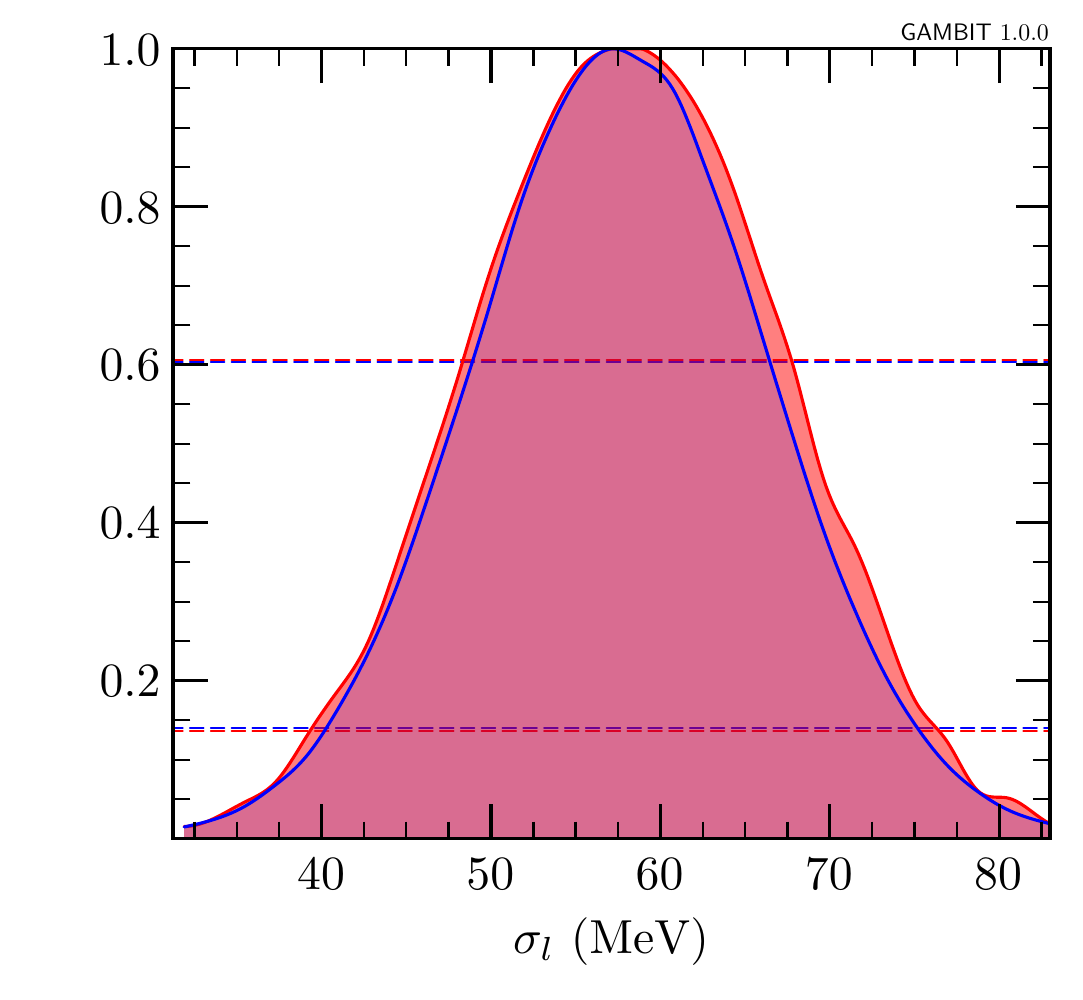}\\
\includegraphics[width=0.212\textwidth]{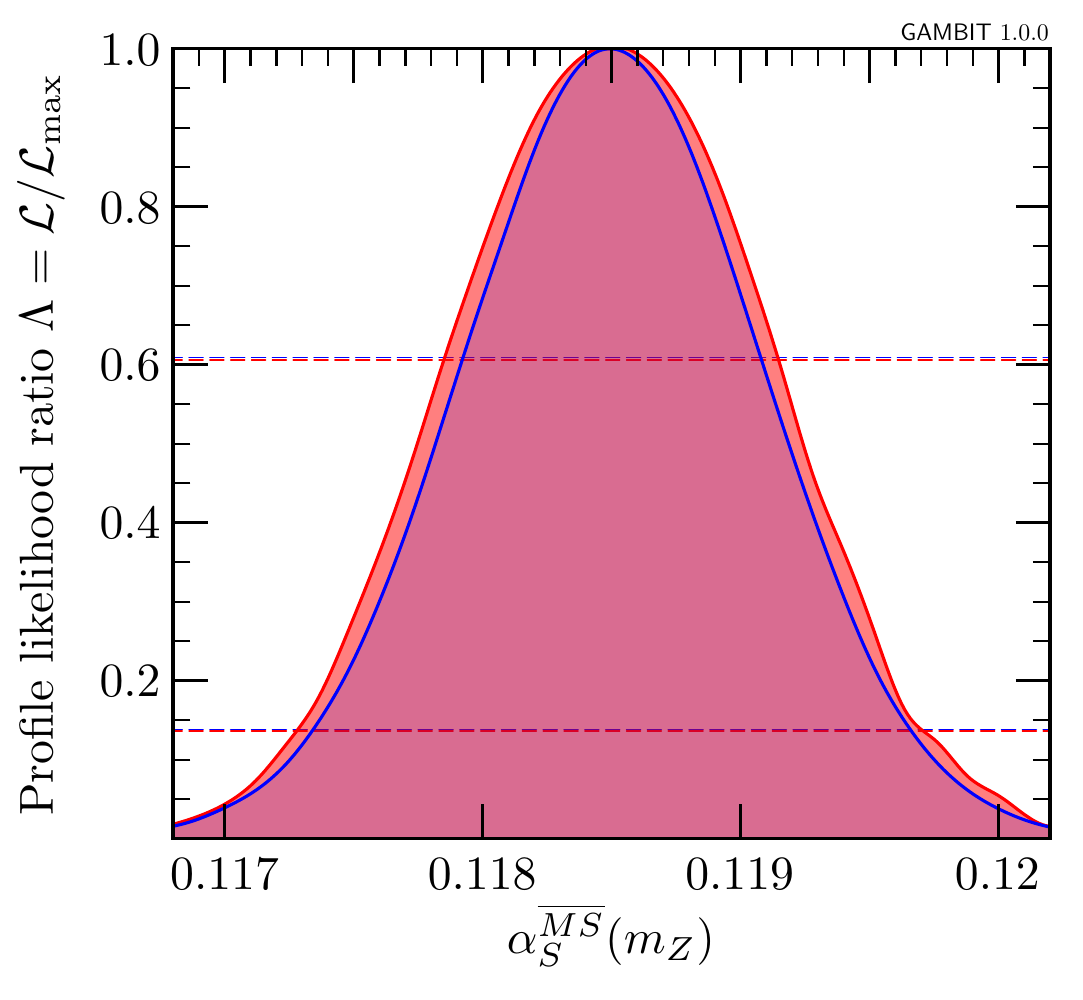}\hspace{-3mm}%
\includegraphics[width=0.212\textwidth]{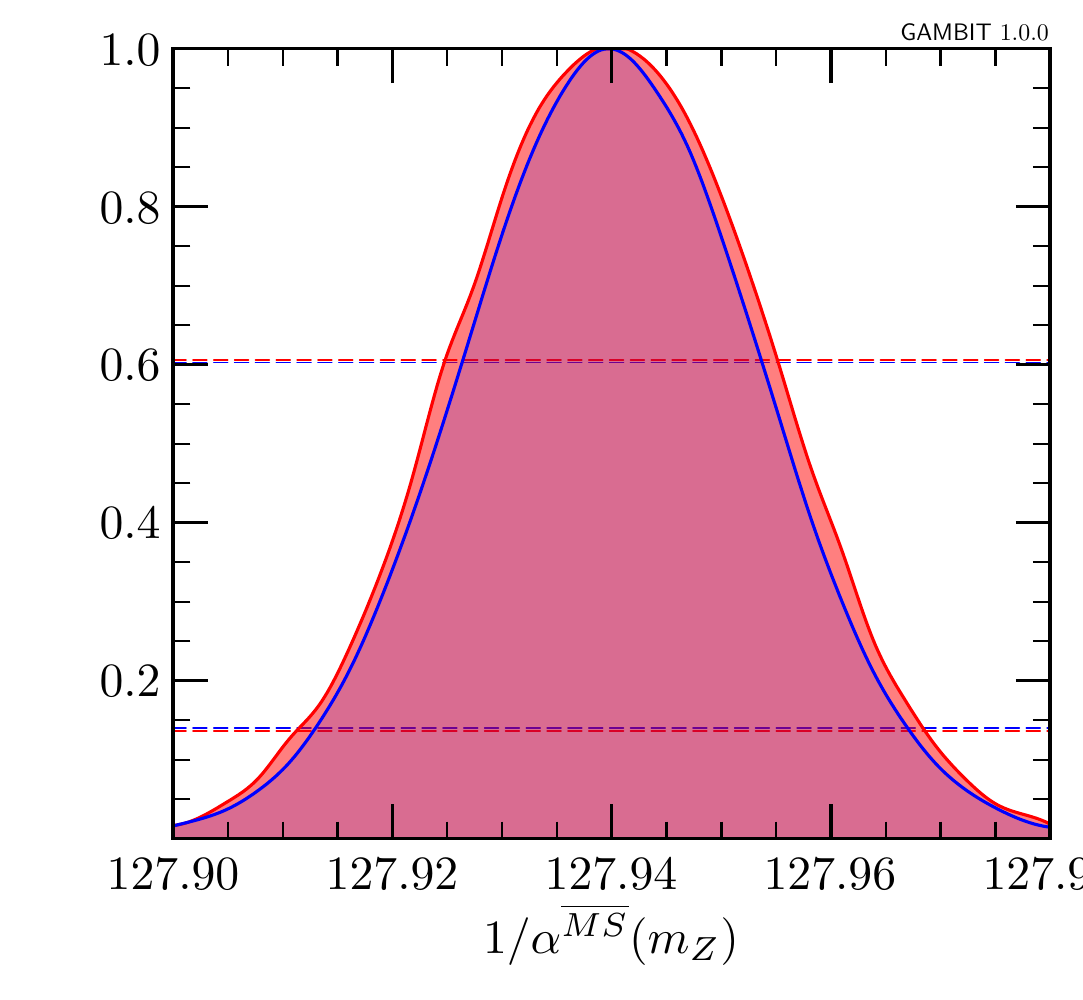}\hspace{-3mm}%
\includegraphics[width=0.212\textwidth]{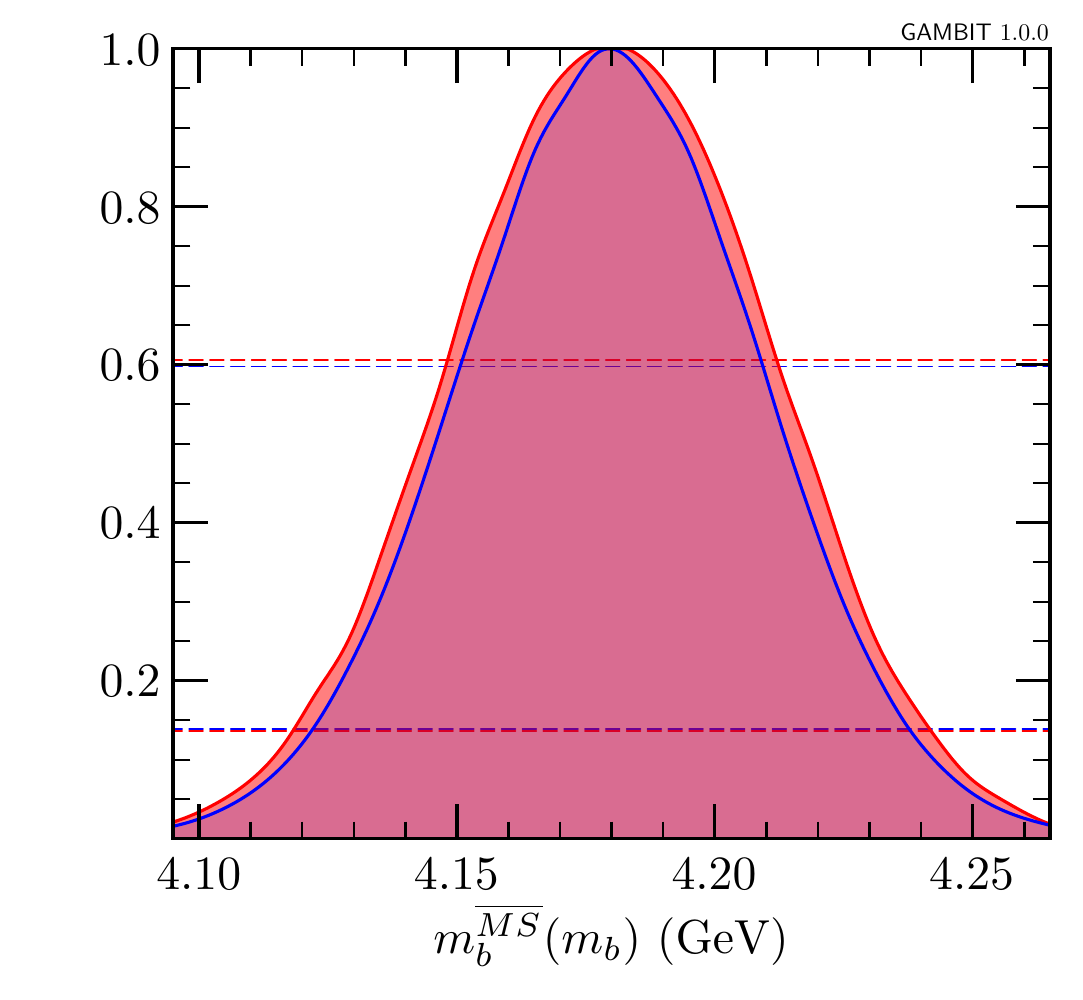}\hspace{-3mm}%
\includegraphics[width=0.212\textwidth]{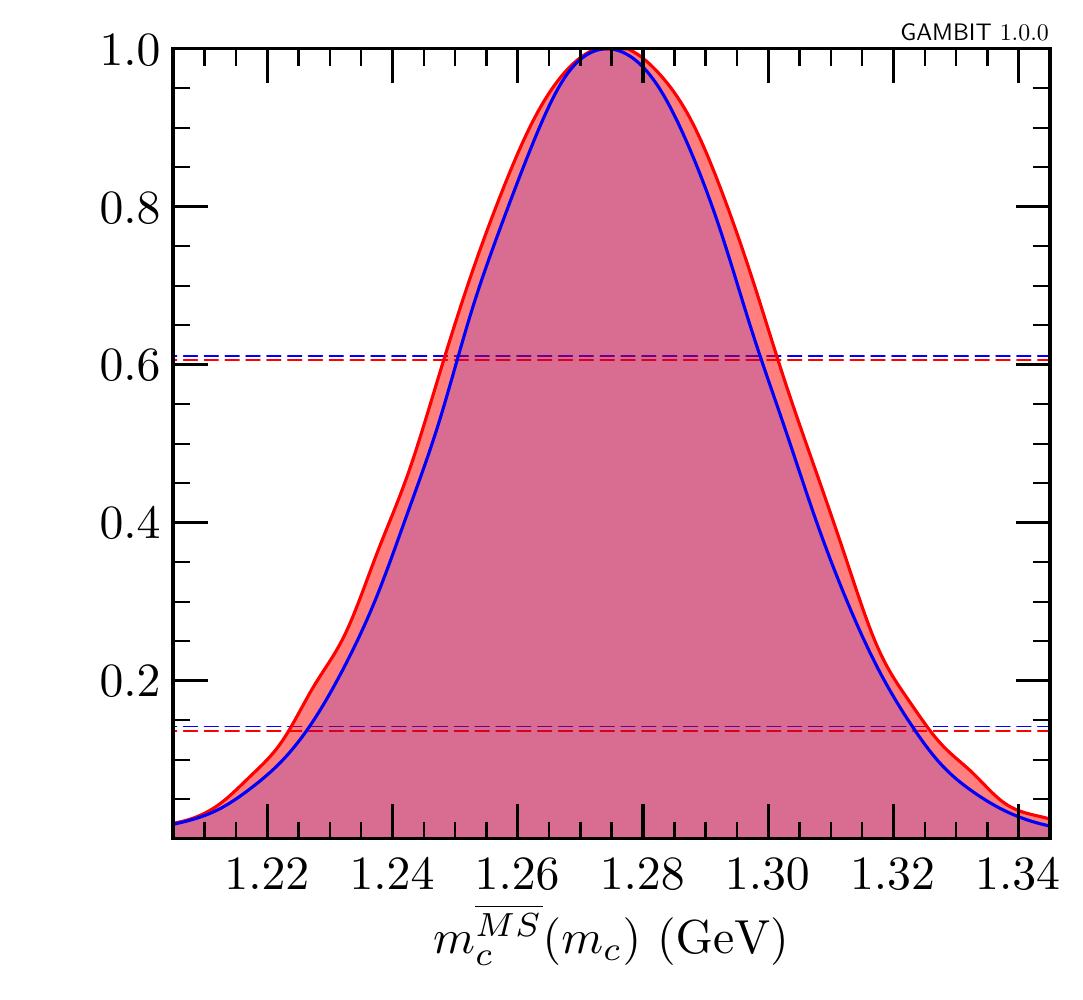}\hspace{-3mm}%
\includegraphics[width=0.212\textwidth]{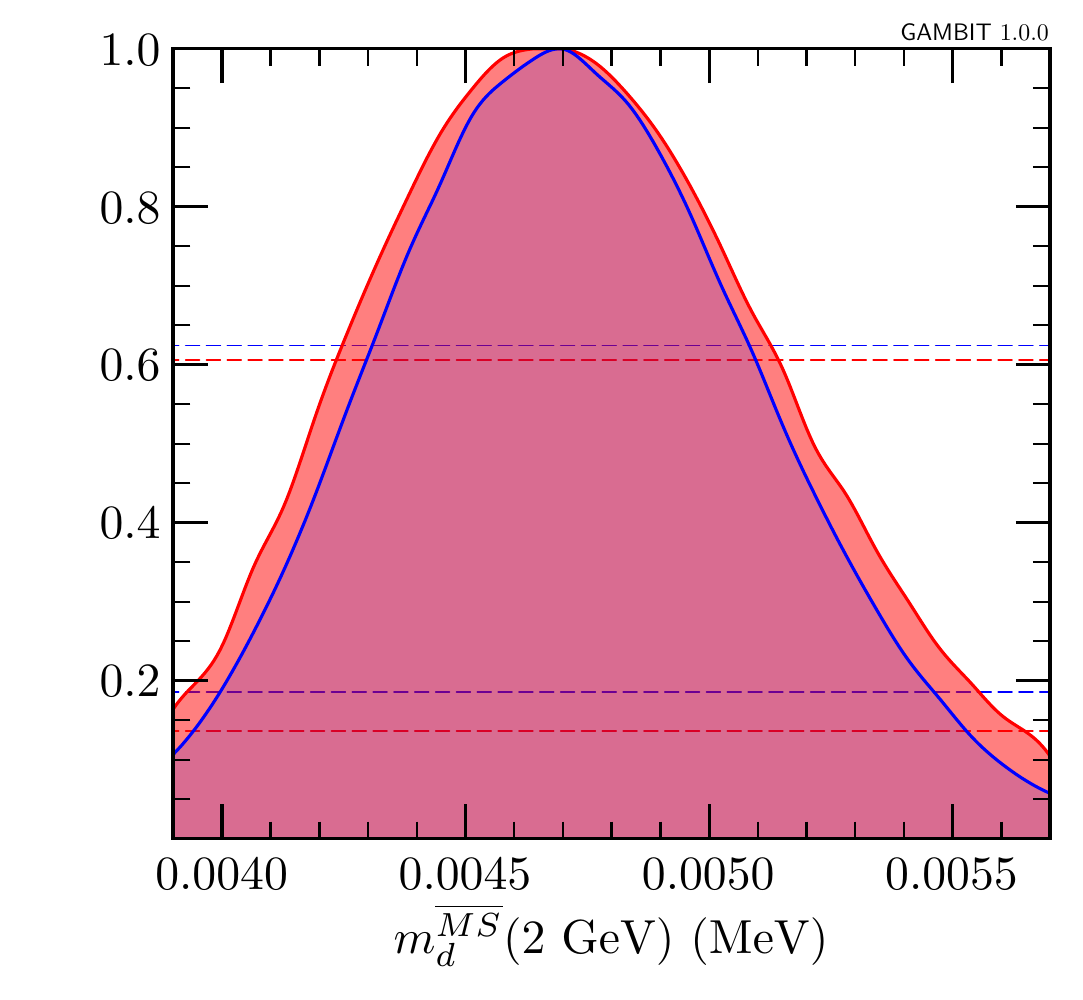}\\
\includegraphics[width=0.212\textwidth]{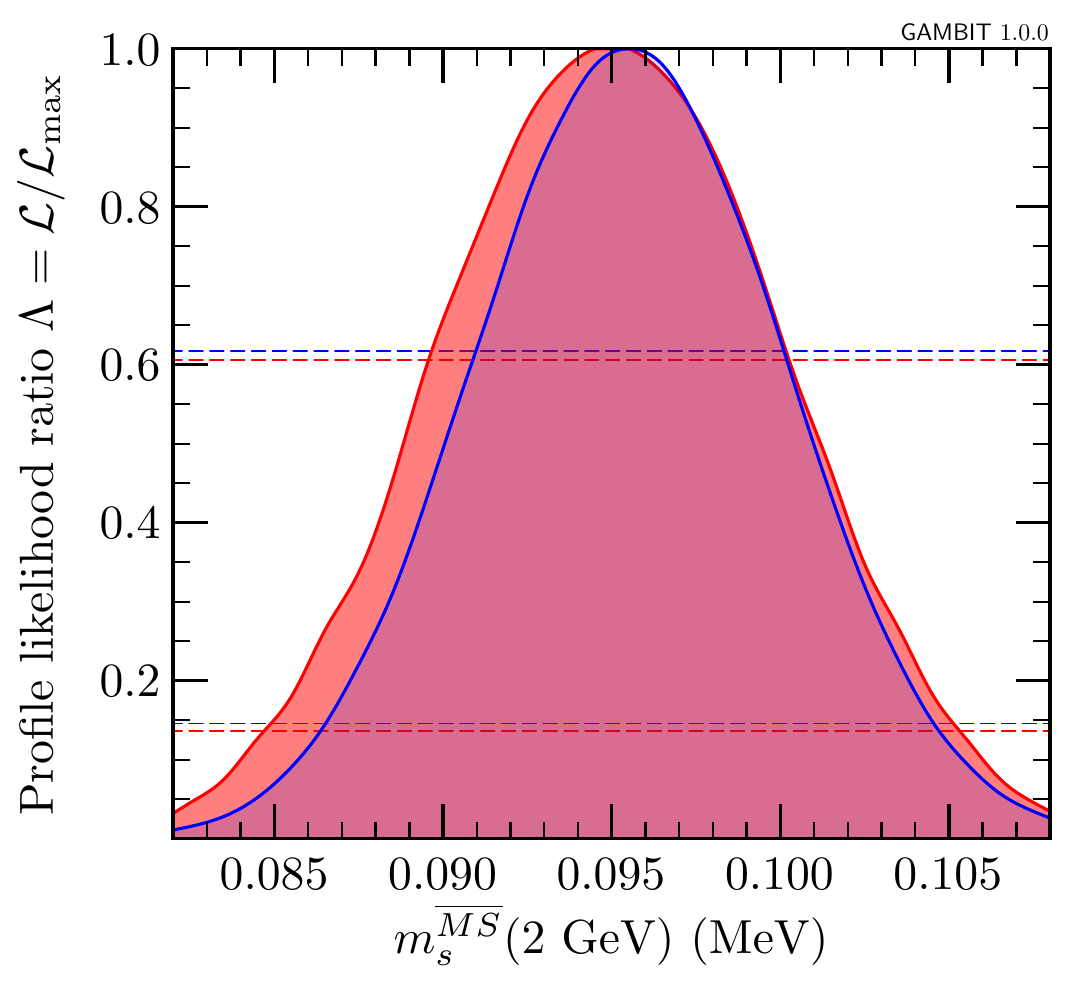}\hspace{-3mm}%
\includegraphics[width=0.212\textwidth]{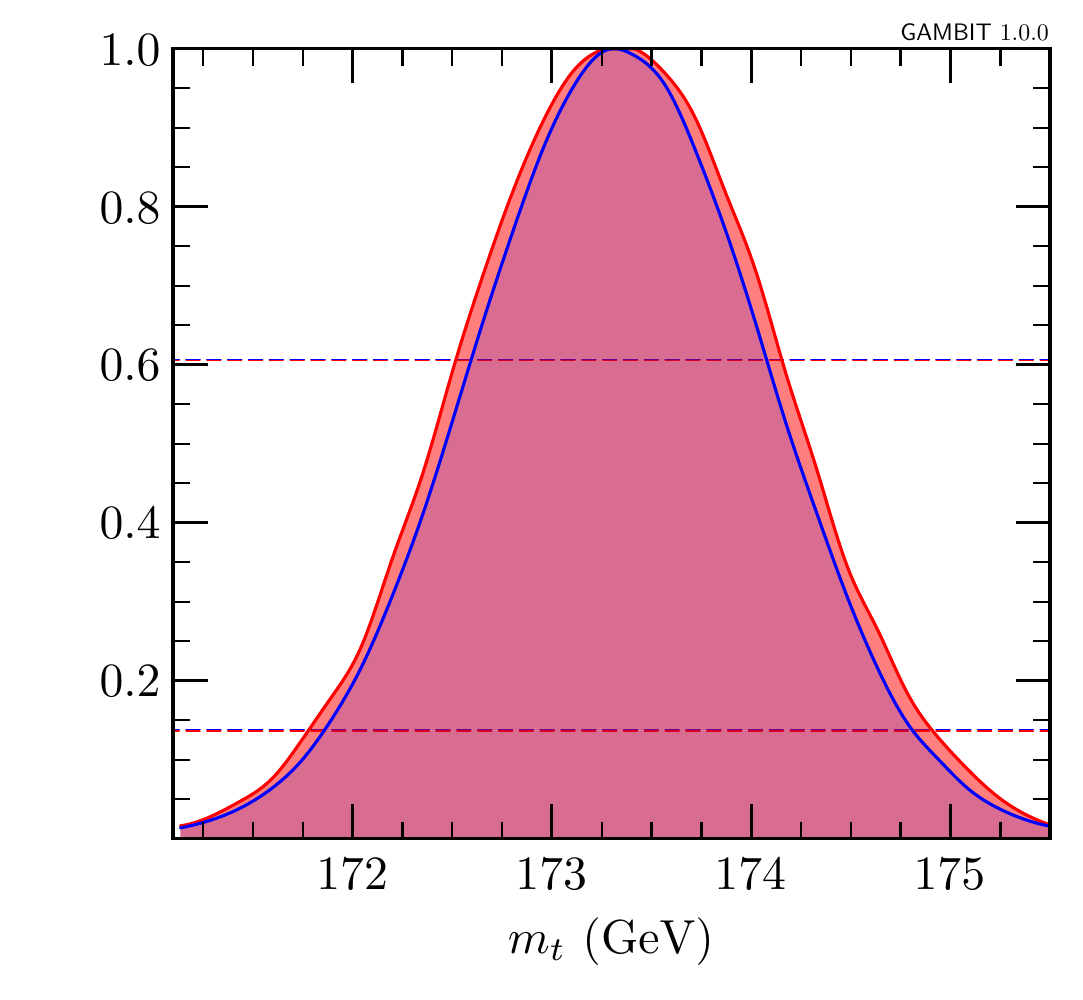}\hspace{-3mm}%
\includegraphics[width=0.212\textwidth]{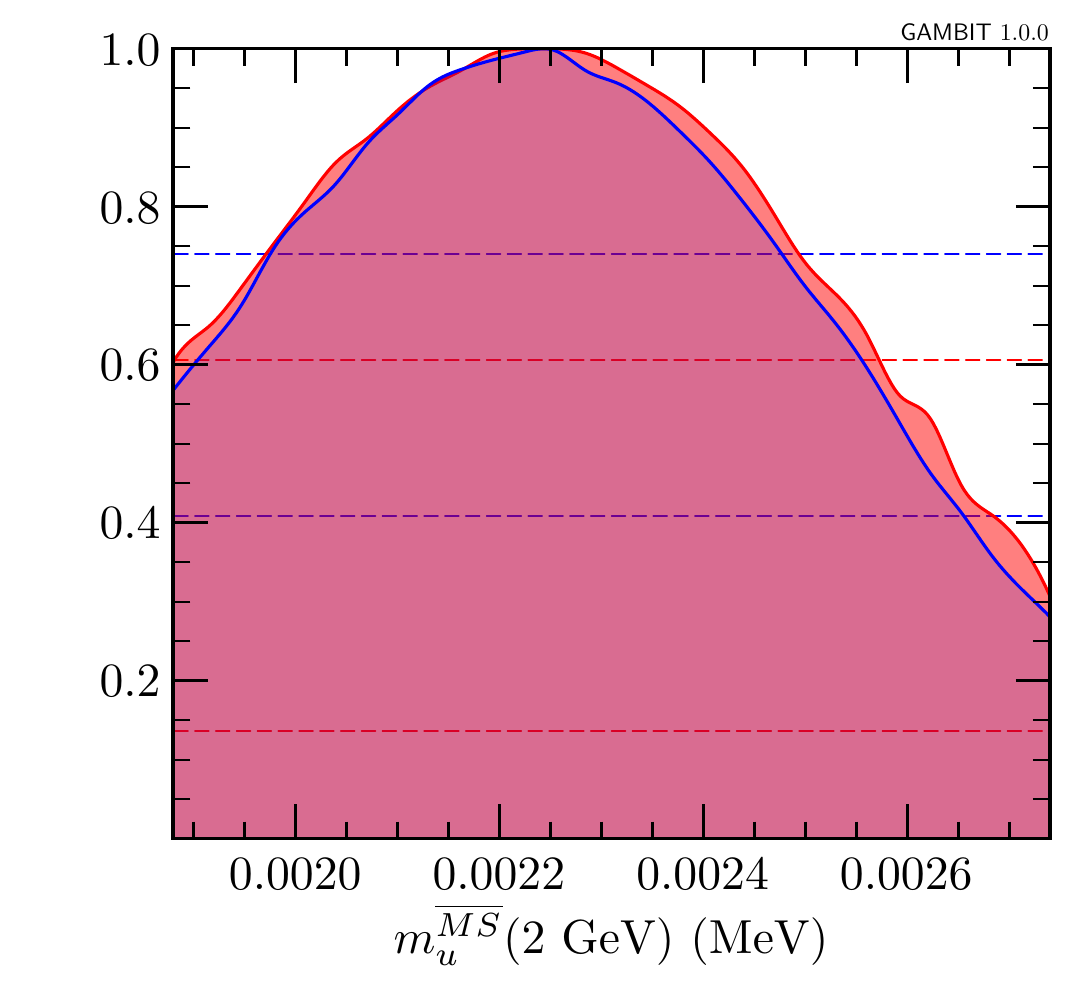}\hspace{-3mm}%
\includegraphics[width=0.212\textwidth]{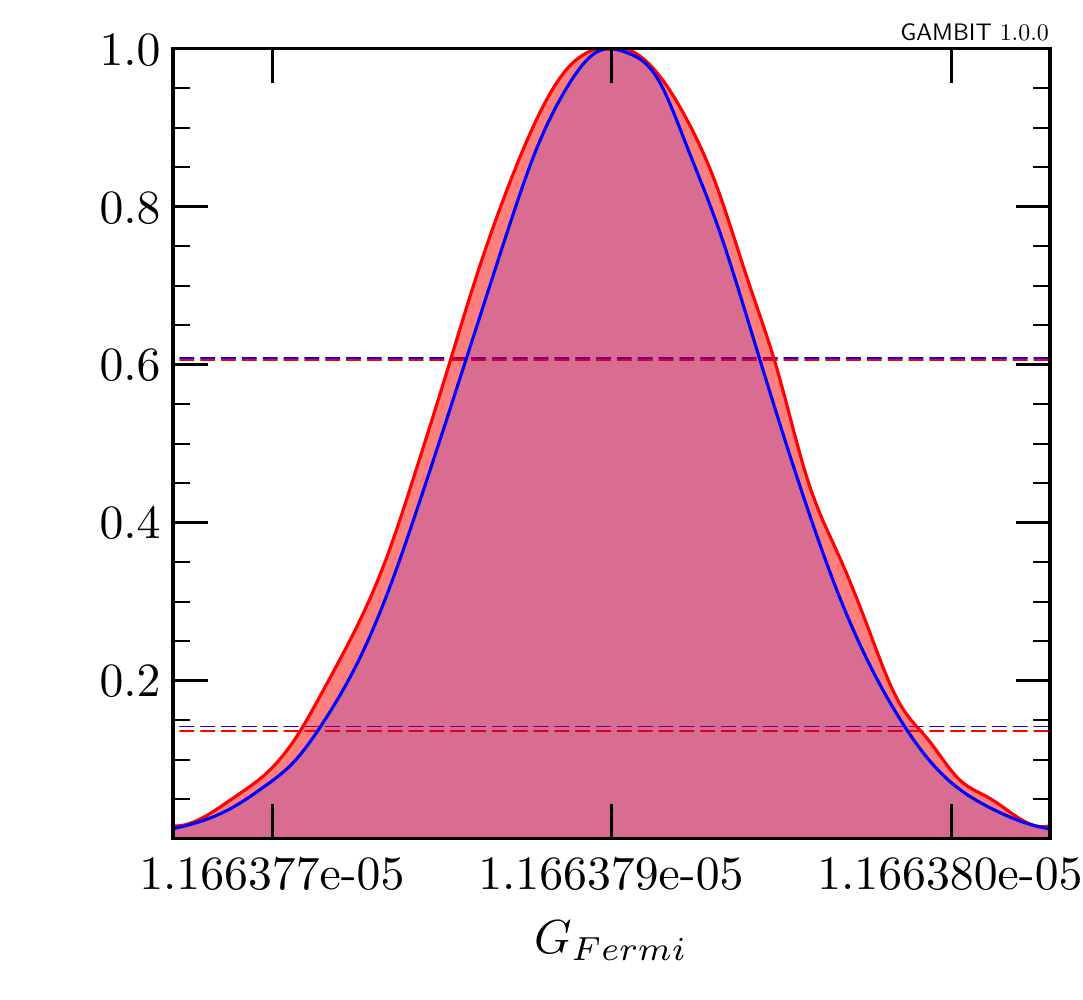}\hspace{-3mm}%
\includegraphics[width=0.212\textwidth]{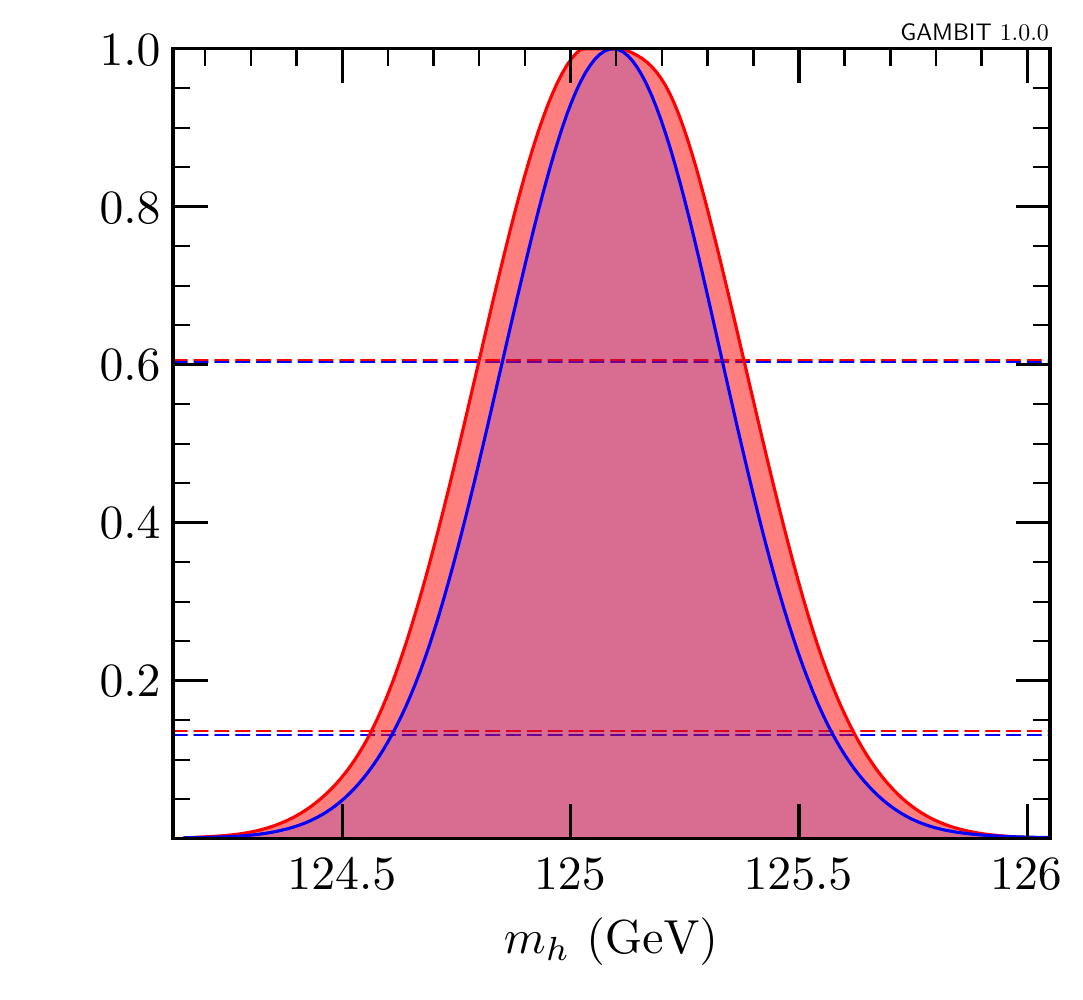}
\caption{One-dimensional profile likelihoods and posterior distributions of the scalar singlet parameters, and all nuisance parameters varied in our fits.  Posterior distributions are shown in blue and profile likelihoods in red.  Dashed lines indicate $1\sigma$ and $2\sigma$ confidence and credible intervals on parameters.}\label{fig::1d_combo}
\end{figure*}

\begin{figure*}[t]
\includegraphics[height=0.85\columnwidth]{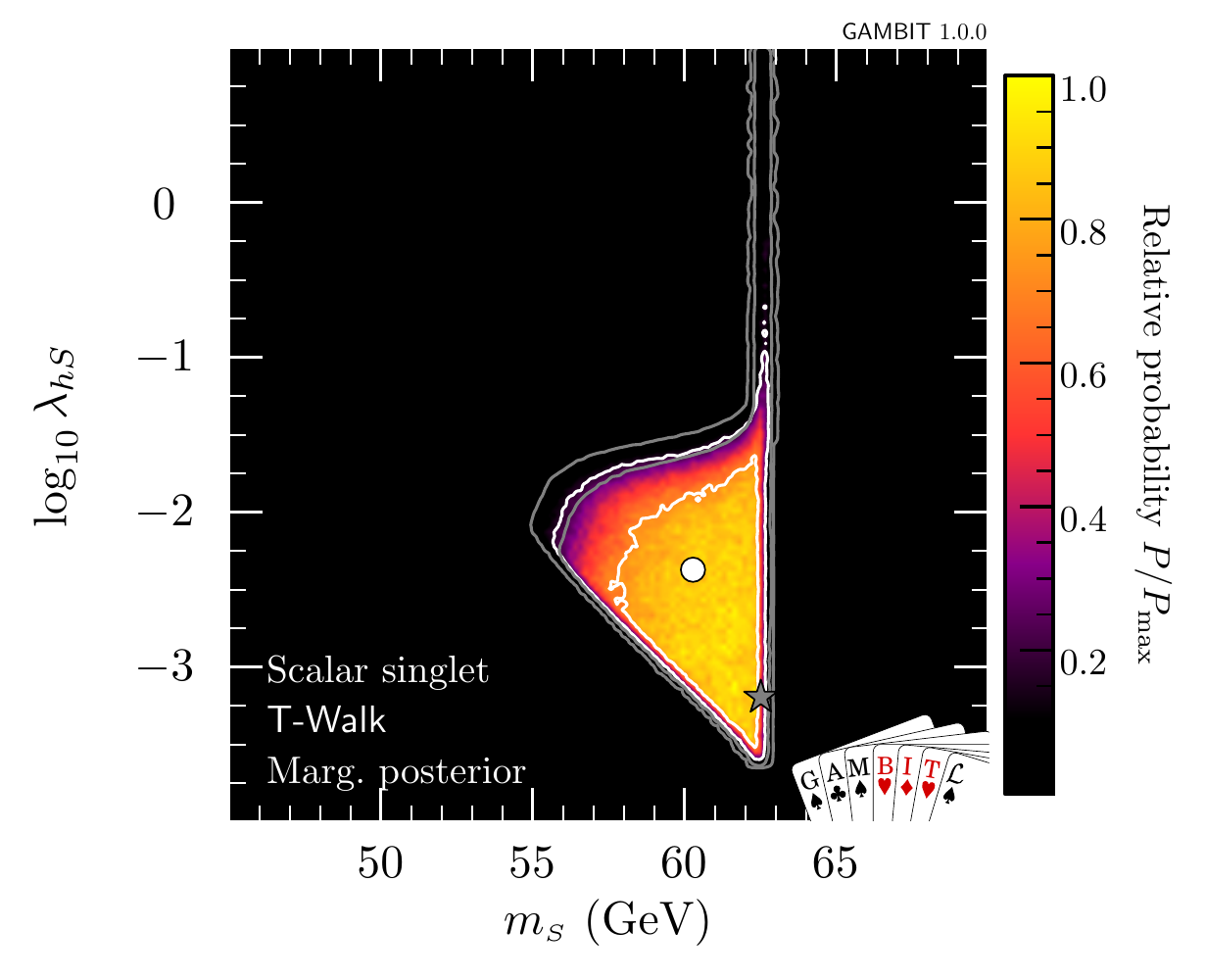}
\includegraphics[height=0.85\columnwidth]{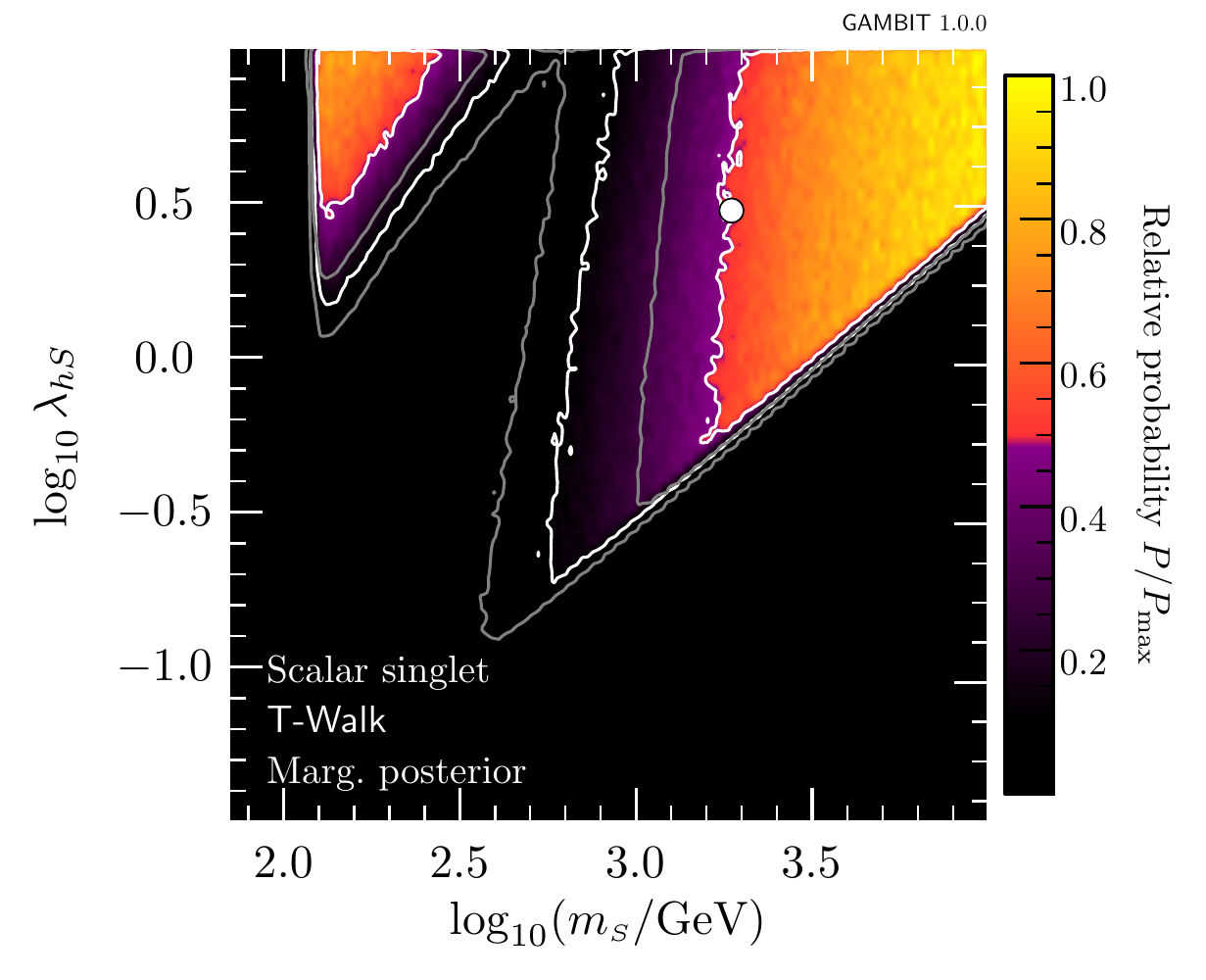}
\caption{Marginalised posterior distributions of the scalar singlet parameters, in low-mass (\textit{left}) and full-range (\textit{right}) scans.  White contours mark out $1\sigma$ and $2\sigma$ credible regions in the posterior.  The posterior mean of each scan is shown as a white circle.  Grey contours show the profile likelihood $1\sigma$ and $2\sigma$ confidence regions, for comparison.  The best-fit (maximum likelihood) point is indicated with a grey star.} \label{fig::2d_post}
\end{figure*}

\section{Results}
\label{sec:results}

\subsection{Profile likelihoods}

Results of our global fit analysis with all nuisances included are presented as 2D profile likelihoods in the singlet parameters in Fig.~\ref{fig::Ms_lhs}, and in terms of some key observables in Figs.~\ref{fig::Ms_oh2} and \ref{fig::Ms_oh2_scaled}.  We also show the one-dimensional profile likelihoods for all parameters in red in Fig.~\ref{fig::1d_combo}.

The viable regions of the parameter space agree well with those identified in the most recent comprehensive studies \cite{Cline13b,Beniwal}.  Two high-mass, high-coupling solutions exist, one strongly threatened from below by direct detection, the other mostly constrained from below by the relic density.  The leading $\lhs^2$-dependence of $\sigma_\text{SI}$ and $\sigma v$ approximately cancel when direct detection signals are rescaled by the predicted relic density, suggesting that the impacts of direct detection should be to simply exclude models below a given mass.  However, the relic density does not scale exactly as $\lhs^{-2}$, owing to its dependence on the freeze-out temperature, resulting in an extension of the sensitivity of direct detection to larger masses than might be na\"ively expected, for sufficiently large values of $\lhs$.\footnote{This point is discussed in further detail in Sect.\ 5 of Ref.\ \cite{Cline13b}.}  This is the reason for the division of the large-mass solution into two sub-regions; at large coupling values, the logarithmic dependence of the relic density on $\lhs$ enables LUX and PandaX to extend their reach up to singlet masses of a few hundred GeV.  This is also slightly enhanced by additional $\lhs^3$ and $\lhs^4$ terms in $\langle \sigma v \rangle_{0,hh}$, which are responsible for the `kink' seen in the border of the grey regions at $\ms\sim600$\,GeV in the left and right panels of Fig.\ \ref{fig::Ms_oh2}.

The resonance region persists, despite being beset from all sides: invisible Higgs from above, relic density from below, indirect detection from higher masses, and direct detection from lower masses.  We find a narrow ``neck'' of degenerate maximum likelihood directly on the resonance, with a best fit located at $\ms=62.51$\,GeV, $\lhs = 6.5\times 10^{-4}$.  The width of this region is set by a number of things:\begin{enumerate}
\item the actual separation between the areas allowed by the invisible width and direct detection constraints, which press in from $\ms<\mh/2$ and $\ms>\mh/2$ respectively,
\item the uncertainty on the Higgs mass, which blurs the exact $\ms$ value of the resonance by $\sim$480\,MeV at the level of the $2\sigma$ contours, and
\item the width of the bins into which we sort samples for plotting, which prevents anything from being resolved on scales below $\Delta\ms\sim170$\,MeV in the left panel of Fig.~\ref{fig::Ms_lhs}.
\end{enumerate}

In addition to correctly identifying the allowed region of the parameter space, we obtain additional information from the global fit analysis beyond that seen from pure exclusion studies.  Using the relic density as an upper limit, all points for which $\Omega_{\sss S}h^2\leq\Omega_\text{DM} h^2$ have a null log-likelihood contribution, and are thus treated equally above the line in parameter space where $\Omega_{\sss S}h^2=0.1188$.  Where $\Omega_{\sss S}h^2<\Omega_\text{DM} h^2$, we rescale the local DM density ($\rho_0$) as well as that in dwarfs $\rho_{\sss S}$, so the direct and indirect detection likelihoods are not flat within the allowed region.  Were we not to rescale signals self-consistently for the predicted relic density, the areas excluded by direct and indirect detection in the first two panels of Fig.~\ref{fig::Ms_oh2} would instead closely track the standard direct and indirect sensitivity curves that many readers will be familiar with.  This can be seen more clearly in Fig.~\ref{fig::Ms_oh2_scaled}, where we plot cross-sections rescaled by the appropriate power of $\Omega_{\sss S} / \Omega_\text{DM}$, together with the experimental constraints from \textit{Fermi}-LAT, LUX and PandaX.

\begin{table}[t!]
\center{
\begin{tabular}{l r r r}
Likelihood contribution & Ideal & Best Fit & $\Delta\ln\mathcal{L}$\\
\hline
Relic density 	&		5.989	& 5.989	& 0 \\
LUX Run I 2015 	&		$-$0.640	&$-$0.640	&0 \\
LUX Run II 2016 	&		$-$1.467	&$-$1.468&	0.001 \\
PandaX 2016  &			$-$1.886	&$-$1.887&	0.001 \\
SuperCDMS 2014 	&		$-$2.248	&$-$2.248&	0 \\
XENON100 2012 &			$-$1.693&	$-$1.693	&0 \\
IceCube 79 		&	0	&0	&0 \\
$\gamma$ rays (\textit{Fermi}-LAT dwarfs) 	&		$-$33.244	&$-$33.349	&0.105 \\
Higgs invisible width 	&		0	&  0&  0 \\
Hadronic elements $\sigma_{s}$, $\sigma_l$ &			$-$6.115	&$-$6.115	&0 \\
Local DM density $\rho_0$ &			1.142	&1.142	&0 \\
$G_{\text{Fermi}}$ 		&	24.92&	24.92	&0 \\
$\alpha_{\text{EM}}$ 	&		3.350	&3.350&	0 \\
$\alpha_{\text{s}}$ 	&		6.500	&6.500	&0 \\
Higgs mass 			&0.508&	0.508&	0 \\
Top quark mass 	&		$-$0.645&	$-$0.645&	0 \\
Bottom quark mass &			2.588&	2.588	&0 \\
Charm quark mass 	&		2.770	&2.770	&0 \\
Light quark masses &			4.844	&4.844	&0 \\
\hline
Total & 			4.673	& 4.566	& 0.107 \\
\end{tabular}
\caption{\label{tab:maxlike} Contributions to the log-likelihood at the best-fit point, compared to an `ideal' case.  The ideal is defined as the central observed value for detections, and the background-only likelihood for exclusions.  Note that each likelihood is dimensionful, so its absolute value is less meaningful than any offset with respect to another point (see Sec.\ 8.3 of Ref.\ \cite{gambit} for more details of the normalisation used). The best-fit point has $\lhs = 6.5\times 10^{-4}$, $\ms = 62.51$\,GeV.}}
\end{table}

Were we to instead restrict our fits to only those models that reproduce \textit{all} of the DM via thermal production to with the \textit{Planck} uncertainties, we would be left with a narrow band along a small number of edges of the allowed regions we have found.  These edges are indicated with orange annotations in Figs.~\ref{fig::Ms_lhs} and \ref{fig::Ms_oh2}.  At high singlet masses, the value of the late-time thermal cross-section (Eq.\ \ref{thermal_av} for $T=0$) corresponding to this strip is equal to the canonical `thermal' scale of 10$^{-26}$\,cm$^3$\,s$^{-1}$.  At low masses, this strip runs along the lower edge of the resonance `triangle' only, as indirect detection rules out models with $\Omega_Sh^2=0.119$ near the vertical edge (at $\ms = 62$\,GeV).

In Fig.\ \ref{fig::Ms_oh2}, we also show in grey the regions corresponding to Higgs-portal couplings above our maximum considered value, $\lhs=10$, in order to give some rough idea of the area of these plots that we have not scanned (and the area that should almost certainly be excluded on perturbativity grounds were we to do so).  We note that at large $\ms$, the highest-likelihood regions are all at quite large coupling values, where the annihilation cross-section is so high, and the resulting relic density is so low, that all direct and indirect signals are essentially absent -- but where perturbativity of the model begins to become an issue.

\subsection{Best-fit point}

Our best-fit point is located within the low-mass resonance region, at $\lhs = 6.5\times 10^{-4}$, $\ms=62.51$\,GeV.  This point has a combined log-likelihood of $\log(\mathcal{L})=4.566$, shown broken down into its various likelihood components in Table~\ref{tab:maxlike}. To put this into context, we also provide the corresponding likelihood components of a hypothetical `ideal' fit, which reproduces positive measurements exactly, and has likelihood equal to the background-only value for those observables with only a limit.  The overall combined ideal likelihood is $\log(\mathcal{L})=4.673$, a difference of $\Delta\ln\mathcal{L}=0.107$ with respect to our best fit.  The best fit above the resonance is at $\lhs=9.9$, $\ms=132.5$\,GeV, with $\log(\mathcal{L})=4.540$, $\Delta\ln\mathcal{L}=0.133$.

Interpreting $\Delta\ln\mathcal{L}$ defined this way is somewhat fraught, as we do not know its distribution under the hypothesis that the best fit is correct.  However, its definition is almost identical to half the ``likelihood $\chi^2$'' of Baker \& Cousins \cite{Baker:1983tu}, which is known to follow a $\chi^2$ distribution in the asymptotic limit.  Our $\Delta\ln\mathcal{L}$ differs from half the likelihood $\chi^2$ only in that some of the components of the ideal likelihood come from the likelihood of a pure-background model, rather than from setting all predictions to their observed values.  Assuming that $2\Delta\ln\mathcal{L}$ follows a $\chi^2$ distribution, estimating the effective number of degrees of freedom would still be difficult, as our likelihoods include many upper limits and Poisson terms, some of which have already been conditioned on the background expectation, and some of which have not.  The difference between the ideal and the best-fit likelihood does nonetheless give some indication of the degree to which the Singlet DM model can simultaneously explain all data in a consistent way, and how much worse it does than the ideal model.  In this sense, it gives information similar in character to the modified $p$-value method known as CLs \cite{CLs1, CLs2, Zech}, which was explicitly designed for excluding models that gave poorer fits than the background model, by conditioning on the background.  Were one to approximate the distribution of $2\Delta\ln\mathcal{L}$ as $\chi^2$ with e.g.\ 1--2 effective degrees of freedom, this would correspond to a rough $p$ value of between 0.6 and 0.9 in both the resonance and the high-mass region -- a perfectly acceptable fit. 

Next we consider parameter combinations where the singlet constitutes the entire observed relic density of DM, by restricting discussion to points with $\Omega_Sh^2$ within $1\sigma$ of the \textit{Planck} value $\Omega_\text{DM}h^2 = 0.1188\pm0.006$ (the uncertainty includes theoretical and observational contributions added in quadrature).  In this case, the best fit occurs at the bottom of the resonance, at $\lhs=2.9 \times 10^{-4}$, $\ms = 62.27$\,GeV.  This point has $\log(\mathcal{L})=4.431$, which translates to $\Delta\ln\mathcal{L}=0.242$ compared the ideal model.  In the high-mass region, the best fit able to reproduce the entire observed relic density is at $\lhs = 3.1$, $\ms = 9.79$\,TeV, and has $\log(\mathcal{L})=4.311$ ($\Delta\ln\mathcal{L}=0.362$).  If we were to approximate the distribution of $2\Delta\ln\mathcal{L}$ as $\chi^2$ with 1--2 degrees of freedom, this would correspond to $p$ values of between 0.5 and 0.8 for the resonance point, and between 0.4 and 0.7 for the high-mass point.  Again, these would suggest that the fit is perfectly reasonable.   This indicates that there is no significant preference from data for scalar singlets to make up either all or only a fraction of the observed DM.

The four best-fit points and the corresponding relic densities are presented in Table \ref{tab:best_fit}.

\begin{table*}[t!]
\center{
\setlength\tabcolsep{2.4pt}
\begin{tabular}{l@{\hspace{4mm}}l@{\hspace{4mm}}l@{\hspace{4mm}}l@{\hspace{4mm}}l@{\hspace{4mm}}l l l l}
  Mode & Statistic & Relic density condition & $\lhs$ & $\ms$ (GeV) &  $\Omega_{\sss S} h^2$  & $\log(\mathcal{L})$ & $\Delta\ln\mathcal{L}$ \\
  \toprule
  Low mass       & Best fit       & $\Omega_{\sss S} h^2\lesssim\Omega_{\sss DM} h^2$       & $\num{6.5e-4}$	& $62.51$	& $0.0179$ & $4.566$ & $0.107$\\
                 & Best fit       & $\Omega_{\sss S} h^2\sim\Omega_{\sss DM} h^2$    & $\num{2.9e-4}$ & $62.27$ & $0.1129$ & $4.431$ & $0.242$\\
                 & Posterior mean & $\Omega_{\sss S} h^2\lesssim\Omega_{\sss DM} h^2$       & $\num{4.3e-3}$ & $60.28$ & & & \\
  \midrule
  High mass      & Best fit       & $\Omega_{\sss S} h^2\lesssim\Omega_{\sss DM} h^2$       & $9.9$ & $132.5$         &  $\num{1.2e-8}$  & $4.540$ & $0.133$\\
                 & Best fit       & $\Omega_{\sss S} h^2\sim\Omega_{\sss DM} h^2$    & $3.1$ & $\num{9.790e3}$ & $0.1131$           & $4.311$ & $0.362$\\
                 & Posterior mean & $\Omega_{\sss S} h^2\lesssim\Omega_{\sss DM} h^2$ & $3.0$ & $1867$ & & & & \\
  \bottomrule
\end{tabular}
\caption{\label{tab:best_fit}Details of the best-fit points and posterior means, differentiated into the two main likelihood modes.  Best fits are given for the case where the singlet relic density is within 1$\sigma$ of its observed value, and for the case where singlet particles may be a sub-dominant component of dark matter.  We omit the values of the 13 nuisance parameters, as they do not deviate significantly from the central values of their associated likelihood functions.}}
\end{table*}

\subsection{Bayesian posteriors}

By using multiple scanning algorithms in our fits, we are also able to consider marginalised posterior distributions for the singlet parameters.  In Figure \ref{fig::1d_combo}, in blue we also plot one-dimensional marginalised posteriors for all parameters, from our full-range posterior scan with the \twalk sampler.\footnote{We choose \twalk for this rather than \multinest, as we find that \multinest biases posteriors towards ellipsoidal shapes; see \cite{scannerbit} for more details and example posterior maps for this same physical model.}  The one-dimensional posterior for $\ms$ shows that although the full-range scan has managed to detect the resonance region, this area has been heavily penalised by its small volume in the final posterior, arising from the volume effect of integrating over nuisance parameters to which points in this region are rather sensitive, such as the mass of the Higgs.  The penalty is sufficiently severe that this region drops outside the $2\sigma$ credible region in the $\ms$-$\lhs$ plane.  We therefore focus only on the high mass modes in the righthand panel of Fig.~\ref{fig::2d_post}, where we show the posterior from the full-range scan.

Because it is restricted to the resonance region, the low-range scan (left panel of Fig.~\ref{fig::2d_post}) shows the expected relative posterior across this region.  The fact that the resonance is so strongly disfavoured in the full-range posterior scan is an indication of its heavy fine-tuning, a property that is naturally penalised in a Bayesian analysis. This mode of the posterior accounts for less than 0.4\% of the total posterior mass, indicating that it is disfavoured at almost $3\sigma$ confidence.

For the sake of understanding the prior dependence of our posteriors, we also carried out a single scan of the full parameter range with flat instead of log priors on $\ms$ and $\lhs$, using \multinest with the same full-range settings as in Table~\ref{table:scanner_params}.  Unsurprisingly, the resulting posterior is strongly driven by this (inappropriate) choice of prior, concentrating all posterior mass into the corner of the parameter space at large $\lhs$ and $\ms$.  The $1\sigma$ region lies above $\lhs \sim 3$, $\ms \sim 3$\,TeV, and the $2\sigma$ region above $\lhs \sim 1$, $\ms \sim 1$\,TeV.

\subsection{Vacuum stability}

Finally, we check vacuum stability for some interesting benchmark points.

So far, our calculations have not required any renormalisation group evolution or explicit computation of pole masses. We have simply taken the tree-level expression for $\ms$ (Eq.\ \ref{m_S_tree}) to indicate the pole mass, and varied it and $\lhs$ as free parameters. To test vacuum stability using \MSbar renormalisation group equations (RGEs), we need to instead use these parameters along with the values of the nuisance parameters to set up boundary conditions for a set of \MSbar RGEs.  We determine values for the \MSbar parameters that give consistent pole masses using \FlexibleSUSY\footnote{\FlexibleSUSY uses numerical routines from \SOFTSUSY \cite{Allanach:2001kg,Allanach:2013kza}.} 1.5.1 \cite{Athron:2014yba}, with \SARAH\ 4.9.1 \cite{Staub:2008uz,Staub:2010jh,Staub:2012pb,Staub:2013tta}. In doing this, it becomes necessary to specify the parameter $\ls$, which we set to zero at the renormalisation scale $m_Z$. \specbit can then evolve the \MSbar parameters to higher scales, using the two-loop RGEs of \FlexibleSUSY, in order to test vacuum stability and also perturbativity.

For our best-fit point, the Higgs-portal coupling $\lhs$ is too small to make a noticeable positive contribution to the running of the Higgs self-coupling, which reaches a minimum value of $-0.0375935$ at $2.523\times10^{17}$\,GeV.  The electroweak vacuum remains
meta-stable for this point, with no substantial change in phenomenology compared to the SM, where for the same Higgs and top quark masses the quartic Higgs coupling has a minimum of $-0.037631$ at $2.514\times10^{17}$\,GeV.

Next we consider a high-mass point within our $1\sigma$ allowed region: $\lhs = 0.5$, $\ms = 1.3$\,TeV.  This point has a large enough coupling $\lhs$ that the minimum quartic Higgs coupling is positive: $0.0522133$ at $1.40006\times 10^9$ GeV.  We see that it is certainly possible to stabilise the electroweak vacuum within the singlet model whilst respecting all current constraints.


\subsection{Comparison to existing results}\label{sec:compare}

The most recent study of the scalar singlet model with a $\mathbb{Z}_2$ symmetry and a wide range of experimental constraints was that of Beniwal et al. \cite{Beniwal}.  This recent study is an ideal candidate with which to compare our results, in order to check for consistency and determine the impacts of the newest experimental constraints.  There are two important differences in the ingredients of our study and that of Beniwal et al.  First, we include stronger DM direct detection constraints from LUX \cite{LUXrun2} and PandaX \cite{PandaX2016}, which exclude a large part of the parameter space.  Second, we scan many relevant nuisance parameters, whereas previous studies have taken them as fixed.  The effect of this can be seen along the boundaries of the confidence intervals, where the viable regions are always at least as large in a scan where the nuisances are allowed to vary as in one where they are fixed.

Considering these differences, we see consistency between the results of this paper and Fig.~4 of Beniwal et al. \cite{Beniwal}, both in the low and high-mass parts of the $\lhs$, $\ms$ parameter space.  The increased size of the allowed region resulting from the variable nuisance parameters is evident along all contour edges.  The behaviour of the stronger direct detection constraint is also visible, in the top left corner of the triangular part of the allowed region in the left panel of Fig.\ \ref{fig::Ms_lhs}, and on the right side of the ``neck''.  In the high-mass area of the parameter space (right panel of Fig.\ \ref{fig::Ms_lhs}), we also see LUX and PandaX cutting a large triangular region into the allowed parameter space, essentially separating the high-mass solutions into two separate likelihood modes.

\section{Conclusion}
\label{sec:conc}

The extension of the Standard Model by a scalar singlet stabilised by a $\mathbb{Z}_2$ symmetry is still a phenomenologically viable dark matter model, whether one demands that the singlet constitutes all of dark matter or not.  However, the parameter space is being continually constrained by experimental dark matter searches.  This is evident in the global fit that we have presented here, combining the latest experimental results and likelihoods to provide the most stringent constraints to date on the parameter space of this model.  Direct detection experiments will fairly soon probe the entire high-mass region of the model, with XENON1T expected to access all but a very small part of each of the high-mass islands \cite{Cline13b}.  The resonance region will prove more difficult, though some hope certainly exists for ton-scale direct detection to improve constraints from the low-$\ms$ direction, and for future colliders focussed on precision Higgs physics to probe the edge of the region at $\lhs\sim0.02$.

We have seen that the best-fit point found in our scan does not have a notable impact on the stability of the electroweak vacuum, due to its rather small value of $\lhs$.  We have shown that larger values of the portal coupling can completely solve the meta-stability of the electroweak vacuum, even whilst satisfying all experimental constraints.  However, the couplings required to do this are not far below where perturbativity starts to become an issue.  Investigating how these competing constraints impact the allowed and preferred regions of the singlet model in a global fit will be one of the aims of our follow-up study.

In this study we have demonstrated some powerful features of the \GB framework.  It is now possible to easily combine likelihoods and observables from \GB and existing packages in a consistent and computationally efficient way.  We have varied 13 nuisance parameters in addition to the two parameters of the scalar singlet model.  We have searched this parameter space using the most modern scanning algorithms available, to provide both frequentist and Bayesian statistical interpretations.  By using parallel computing resources, we have achieved this with, for example, a maximum runtime for any \diver scan presented in this paper of just 3\,hr.  Finally, due to the modularity and flexibility of the \GB system, it will be possible to include new likelihoods and/or change parts of the calculation at any time in future, in order to quickly update the analysis to take into account new experimental developments.

All input files, samples and best-fit benchmarks produced for this paper are publicly accessible from \textsf{Zenodo} \cite{the_gambit_collaboration_2017_801511}.  The \GB software is available from \href{http://gambit.hepforge.org}{gambit.hepforge.org}.

\begin{acknowledgements}
We thank Jim Cline, Andrew Fowlie, Tom\'as Gonzalo, Julia Harz, Sebastian Hoof, Kimmo Kainulainen, Roberto Ruiz, Roberto Trotta and Sebastian Wild for useful discussions, and Lucien Boland, Sean Crosby and Goncalo Borges of the Australian Centre of Excellence for Particle Physics at the Terascale for computing assistance and resources.  \gambitacknos
\end{acknowledgements}

\bibliography{R1}

\end{document}